\documentclass[11pt,a4paper]{article}
\usepackage{amsfonts,amssymb,epsfig,amsmath,enumitem}
\usepackage[utf8]{inputenc}
\usepackage{textcomp}
\usepackage{caption,subcaption}
\usepackage{xcolor}

\renewcommand{\thefootnote}{\fnsymbol{footnote}}

\allowdisplaybreaks

\begin{document}

\makeatletter \@addtoreset{equation}{section} \makeatother
\renewcommand{\theequation}{\thesection.\arabic{equation}}
\renewcommand{\thefootnote}{\alph{footnote}}

\begin{titlepage}
\begin{center}

\hfill {\tt KIAS-P15053}\\
\hfill {\tt SNUTP15-010}\\

\vspace{2cm}

{\Large\bf Higgsing towards E-strings}

\vspace{2cm}

{\large Joonho Kim$^1$, Seok Kim$^2$, Kimyeong Lee$^1$}

\vspace{0.7cm}

\textit{$^1$School of Physics, Korea Institute for Advanced Study,
Seoul 130-722, Korea.}\\

\vspace{0.2cm}

\textit{$^2$Department of Physics and Astronomy \& Center for
Theoretical Physics,\\
Seoul National University, Seoul 151-747, Korea.}\\

\vspace{0.7cm}

E-mails: {\tt joonhokim@kias.re.kr, skim@phya.snu.ac.kr, klee@kias.re.kr}

\end{center}

\vspace{1cm}

\begin{abstract}

We explore 6d $(1,0)$ superconformal field theories with $SU(3)$ and $SU(2)$ gauge symmetries which cascade after Higgsing to the E-string theory on a single M5 near an   $E_8$ wall.
Specifically, we study the 2d $\mathcal{N}=(0,4)$ gauge theories which   describe   self-dual strings of these 6d theories. The self-dual strings can be also viewed as instanton string solitons of 6d Yang-Mills theories.  We find the 2d anomaly-free gauge theories for self-dual strings, amending the naive ADHM gauge theories which are anomalous, and calculate their elliptic genera.
While these 2d theories respect the flavor symmetry of each 6d SCFT only partially,  their elliptic genera manifest the symmetry fully as these  functions as BPS index are invariant in strongly coupled IR limit.  
Our consistent 2d $(0,4)$ gauge theories also provide new insights on the non-linear sigma models for the instanton strings,
providing novel UV completions of the small instanton singularities.
Finally, we construct new 2d quiver gauge theories for the self-dual strings in 6d E-string theory for multiple M5-branes probing the $E_8$ wall, and find their fully refined elliptic genera.

\end{abstract}

\end{titlepage}

\tableofcontents

\section{Introduction} 
\label{sec:intro}

There are many 6d superconformal field theories (SCFTs) engineered from string theory.
The first examples were the maximally supersymmetric $\mathcal{N}=(2,0)$ SCFTs of $ADE$ types, discovered in \cite{Witten:1995zh} from considering IIB string theory compactified on K3.
The $A_N$-type theory governs the low energy dynamics of parallel and overlapping $(N+1)$ M5-branes \cite{Strominger:1995ac}.
Vast number of $\mathcal{N}=(1,0)$ SCFTs were found from F-theory compactified on elliptic Calabi-Yau 3-folds as well as from branes at low energy
\cite{Morrison:1996na,Seiberg:1996vs,Witten:1996qb,Morrison:1996pp,Blum:1997mm,Brunner:1997gf,Hanany:1997gh,DelZotto:2014hpa}. The classification of 6d $(1,0)$ SCFTs   has been recently explored   in \cite{Morrison:2012np,Heckman:2013pva,Heckman:2015bfa,Bhardwaj:2015xxa}.

The tensor branch of each 6d $(1,0)$ SCFT can be described in terms of tensor multiplets, hypermultiplets, and 
vector multiplets. When there exists a gauge symmetry, the gauge anomaly should vanish. 
Each tensor multiplet can have self-dual strings. They are the sources for the 2-form tensor field $B$ of the tensor multiplet, whose field strength $H$ respects the self-duality $H = \ast_6 H$. Their tensions are proportional to the VEV of a tensor multiplet scalar.  
One interesting example is the self-dual strings of the 6d $(2,0)$ $A_N$ SCFT, 
which can be described by M2-branes connecting any two M5-branes of $(N+1)$ M5-branes \cite{Strominger:1995ac}. 
They are the so-called `M-strings' which have been studied in various literatures such as \cite{Klemm:1996bj,Schwarz:1996pi,Haghighat:2013fc,Haghighat:2013fz}. 
Another typical example is the self-dual strings of the 6d $(1,0)$ SCFT with $E_8$ global symmetry, being induced from M2-branes connecting M5-branes to an $E_8$ boundary wall of M-theory on $S^1/\mathbb{Z}_2$ \cite{Ganor:1996mu,Seiberg:1996vs}. These so-called `E-strings' have been studied in, e.g., \cite{Klemm:1996dt,Minahan:1998vr,Eguchi:2002fc,Kim:2014dza}. We shall often use the terminologies `M-string theory' and `E-string theory' to refer the $(2,0)$ $A_N$ SCFT and the $(1,0)$ SCFT with $E_8$ flavor symmetry, respectively.   

There are a lot of  6d $(1,0)$ SCFTs with gauge symmetries. When a gauge group $G$ is semi-simple, \cite{Danielsson:1997kt,Bershadsky:1997sb} search for all possible gauge theories being free of gauge anomaly. If one considers some low-rank simple groups, i.e., $G = SU(2)$ and $SU(3)$, the allowed matter contents are given by
\begin{alignat}{3}
  SU(3):&\ N_f=0,6,12 \quad\quad {\rm and}\quad\quad& SU(2):&\ N_f=4,10.
\end{alignat}
Apart from the minimal $SU(3)$ case with $N_f=0$, the above theories are obtained by enhancing M-string and E-string theories with gauge symmetries and extra matters. 
One can consult \cite{Bershadsky:1997sb,Heckman:2015bfa} for further possible enhancements of M-string and E-string theories found from F-theory construction. In this paper, we shall focus on the following sequences of 6d SCFTs:  
\begin{alignat}{3}
  \label{eq:higgsing-sequence-m}
  &\quad (SU(3), N_f = 6) &\rightarrow&\ \ (SU(2), N_f = 4) &\rightarrow&\ \ (\text{M-string theory}),\\
  \label{eq:higgsing-sequence-e}  
  &\quad (SU(3), N_f = 12)\ &\rightarrow&\ \ (SU(2), N_f = 10)\ &\rightarrow&\ \ (\text{E-string theory}), 
\end{alignat}
which are connected by Higgs mechanism.
Our main interest in this work is to study the physics of self-dual strings in these theories by constructing their 2d gauge theory description.
For each sequence of 6d SCFTs, we shall find a class of 2d gauge theories which provides a uniform description of self-dual strings.
``Decorations'' of the basic 6d SCFTs and self-dual strings, such as M-strings and E-strings, by 6d gauge symmetries and matters are important to fully understand the recent classification \cite{Heckman:2015bfa} better.
And also, we shall discover new gauge theory description of E-strings themselves, which will exhibit some strong coupling properties of these strings more transparently than \cite{Gadde:2015tra}. 

Recall that E-string theory is engineered from M5-branes probing an $E_8$ wall, which is also called   M9-brane. The M9-brane worldvolume has 4 transverse directions to M5-branes, rotated by
the $SO(4)$ symmetry which can be decomposed into $SU(2)_L \times SU(2)_R$.
While all $SU(2)_L$ charged states decouple from E-string theory on a single M5-brane, the $SU(2)_L$ refinement turns out crucial in case of multiple M5-branes.
Our new description of E-strings fully realizes both $SU(2)_L$ and $SU(2)_R$ in contrast to \cite{Gadde:2015tra} which only see $SU(2)_D \subset SU(2)_L \times SU(2)_R$, enabling us to study the refined spectrum of multiple M5-branes probing an M9-brane in Section~\ref{sec:multiple-m5-branes}.

Let us explain a useful bottom-up approach for finding the 2d gauge theories for self-dual strings. When a 6d SCFT has a gauge symmetry, the equation of motion for $B$ is given by \cite{Berkooz:1996iz,Seiberg:1996qx}
\begin{align}
    \label{eq:tensor-vector-eom}
    d\ast H = d H = \sqrt{c} \, \text{tr} (F \wedge  F),
\end{align}
where $F$ is the field strength of a gauge field, and $c$ is a positive constant which depends on a theory \cite{Seiberg:1996qx}. Self-dual strings are therefore regarded as instanton soliton strings in effective Yang-Mills description of 6d SCFTs in the tensor branch. These strings carry nonzero integer charges of
\begin{align}
  \label{eq:instanton-charge}
    k = \frac{1}{8\pi^2} \int d^4 x\  \text{tr}\, (F \wedge F) \in \mathbf{Z}.
\end{align}
We shall consider self-dual instantons with $k>0$ in this paper.

This integral is taken over the $\mathbf{R}^4$ space transverse to the string worldsheet. If one applies the moduli space approximation to describe the low energy excitations on these strings, the dynamics of self-dual strings are described by non-linear sigma models, whose target space is given by the instanton moduli space $\mathcal{M}$. The space $\mathcal{M}$ has singular locus at which the instanton size shrinks to zero.
This small instanton singularity reflects the UV incompleteness of the 6d effective Yang-Mills theory.
Although it is difficult to deal with the singular target space, the ADHM construction \cite{Atiyah:1978ri} often tells us the sensible UV completions of the non-linear sigma models.
With the guidance from the ADHM construction, one may obtain 2d $\mathcal{N}=(0,4)$ ADHM gauge theories which are weakly coupled in the UV regime. The instanton moduli space $\mathcal{M}$ reappears in the Higgs branch of the ADHM gauge theory \cite{Witten:1994tz} away from the points where the instanton scale becomes zero.

One main problem that we discuss in Section~\ref{sec:strings-from-branes} is that the ADHM construction sometimes leads us to 2d gauge theories that suffer from gauge anomaly. While there is no such issue for self-dual strings in  the sequence to M-strings \eqref{eq:higgsing-sequence-m}, this happens when we consider 6d SCFTs in the Higgsing chain \eqref{eq:higgsing-sequence-e} to E-string theory. 
We find anomaly-free gauge theories by suitably modifying the naive ADHM gauge theories, 
which include the same 6d instanton zero modes.
These anomaly-free theories have the desired non-linear sigma models on $\mathcal{M}$ away from the small instanton singularity, but provide a UV completion at the singularity.
One important feature is that some global symmetries which we expect on the self-dual strings are partly broken in these gauge theories. Namely, the 6d symmetries
%
\begin{align}
    G = SU(3): \ SU(12), \quad\quad\quad G = SU(2): \ SO(20),\quad\quad\quad \text{E-string theory} : \ E_8 ,
\end{align}
  are sometimes partially present in the 2d gauge theories.
The elliptic genera of these consistent 2d theories capture the BPS excitations on self-dual strings and remain valid in the strong coupling low energy limit where the 6d global symmetries should get manifest. The calculation of the elliptic genera and showing   the global symmetry enhancement in the index functions are    done in Sections~\ref{sec:IR-dynamics-symmetry}~and~\ref{sec:multiple-m5-branes}.

The outline of this paper is as follows. In Section~\ref{sec:strings-from-branes}, we construct 2d anomaly-free gauge theories for self-dual strings in 6d SCFTs listed at \eqref{eq:higgsing-sequence-m} and \eqref{eq:higgsing-sequence-e}. We check that they show the IR symmetry enhancements by computing the gauge theory elliptic genera in Section~\ref{sec:IR-dynamics-symmetry}. Section~\ref{sec:multiple-m5-branes} is devoted to study E-string theories with higher dimensional tensor branches, i.e. for multiple M5-branes probing an M9-brane. We obtain the fully flavored spectrum which refines the result of \cite{Gadde:2015tra}. Concluding remarks are given in Section~\ref{sec:conclusion}.

\section{Gauge theories on self-dual strings}
\label{sec:strings-from-branes}

In this section, we construct the anomaly-free $(0,4)$ gauge theories for $k$ self-dual strings in the following sequences of 6d gauge theories:
\begin{alignat}{3}
  \label{eq:higgsing-sequence-m-ch2}
   &\quad (SU(3), N_f = 6) &\rightarrow&\ \ (SU(2), N_f = 4) &\rightarrow&\ \ (\text{M-string theory}),\\
  \label{eq:higgsing-sequence-e-ch2}  
   &\quad (SU(3), N_f = 12)\ &\rightarrow&\ \ (SU(2), N_f = 10)\ &\rightarrow&\ \ (\text{E-string theory}).
\end{alignat}
It is tightly related to finding a suitable UV completion of the non-linear sigma model with the singular target space. As this procedure has not been clearly understood, we shall refer to the string theory realization of 6d SCFTs and self-dual strings as our guide for a possible UV completion.

\begin{figure}[ht]
    \centering\vspace{0.3cm}
    \subcaptionbox{}{
        \includegraphics[width=8cm]{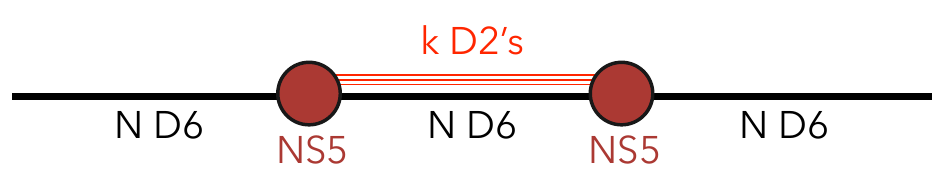}}
    \hfill
    \subcaptionbox{}{
    \begin{tabular}{c|cccccccccc}
      & 0 & 1 & 2 & 3 & 4 & 5 & 6 & 7 & 8 & 9 \\ \hline
      NS5 & $\bullet$ & $\bullet$ & $\bullet$ & $\bullet$ & $\bullet$ & $\bullet$ & $-$ & $-$ &$-$ & $-$\\
      D6 & $\bullet$ & $\bullet$ & $\bullet$ & $\bullet$ & $\bullet$ & $\bullet$ & $\bullet$ & $-$ & $-$ & $-$\\
      D2 & $\bullet$ & $\bullet$ & $-$ & $-$ & $-$ & $-$ & $  \bullet $ & $-$ & $-$ & $-$
    \end{tabular}
    }
    \caption{IIA brane system for 6d $SU(N)$ theory Higgsable to M-string theory}
    \label{fig:m-string-higgsing-sequence}
\end{figure}

The Higgsable theories in \eqref{eq:higgsing-sequence-m-ch2}   are readily realized in type IIA string theory. Suppose that there exist $N$ overlapping D6-branes along the $0123456$ directions. Two parallel  NS5-branes along the $012345$ directions are located on these D6 branes and separated in the $6$ direction. There are $k$ overlapping D2-branes along the $016$ directions connecting two NS5-branes.  These branes are mutually BPS and so preserve some supersymmetry. Figure~\ref{fig:m-string-higgsing-sequence} shows the brane set-up.   

The 7d $U(N)$ SUSY gauge theory on $N$ D6-branes implies that the low energy physics between two NS5-branes is described by the 6d $(1,0)$ $SU(N)$ gauge theory with $2N$ fundamental hypermultiplets, which is coupled to the tensor multiplet from the relative motion of two NS5-branes. The reduction mechanism of the 6d gauge group from $U(N)$ to $SU(N)$ is explained in \cite{Douglas:1996sw,Berkooz:1996iz,Hanany:1997gh}.
The 6d $U(1)$ gauge symmetry with charged matters was anomalous, as only hypermultiplets could contribute to the $U(1)$ gauge anomaly. One can however obtain a sensible theory through the `linear hypermultiplet' having four scalars $\xi^{1,2,3}$ and $\theta$ , which correspond to the position coordinates of NS5-branes along $789$ and M-circle directions. The   $\xi^{1,2,3}$ appear as parameters in the D-term  and the $\theta$ appears as $\theta \,{\rm tr} F^3$.
The kinetic terms for linear hypermultiplet scalars are given by \cite{Douglas:1996sw}
\begin{align}
    (\partial_\mu \xi^A)^2 + (\partial_\mu \theta -   A_\mu)^2
\end{align}
where the $U(1)$ gauge invariance requires $\theta$ to transform as $\theta  \rightarrow \theta +     \epsilon$ under $A_\mu \rightarrow A_\mu + \partial_\mu \epsilon$.
This shows that the $U(1)$ gauge field is massive and the $U(1)$ gauge symmetry is spontaneously broken.   The low energy physics is therefore governed by the $SU(N)$ gauge theory with $2N$ fundamental hypermultiplets. This 6d theory is  anomaly-free.

\begin{figure}[h!]
    \centering\vspace{0.3cm}
    \subcaptionbox{}{
        \includegraphics[height=2.8cm]{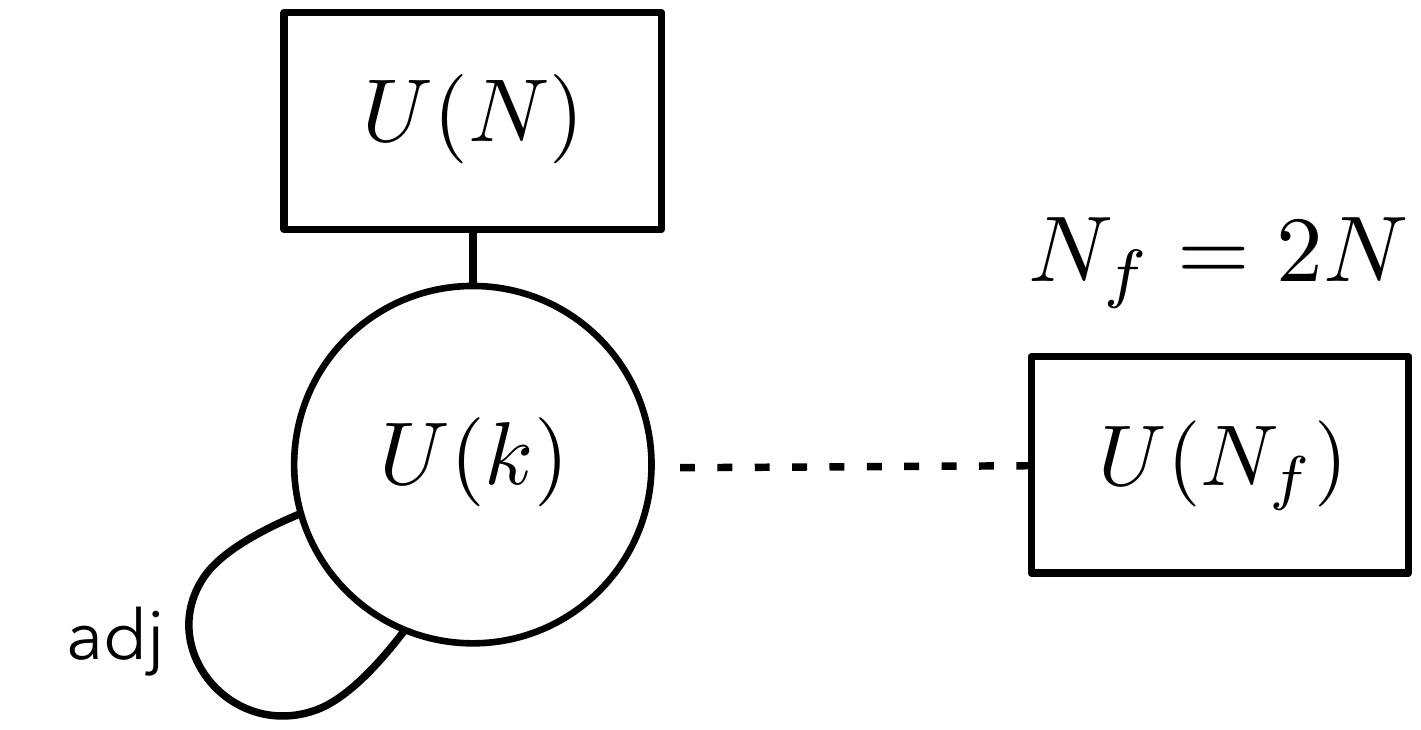}
    }\hfill
    \subcaptionbox{}{
      \begin{tabular}{ c | c | ccc}
        Field & Type & $U(k)$ & $U(N)$ & $U(N_f)$\\
        \hline
        $(A_\mu, \lambda^{A\dot{\alpha}})$ & vector & \textbf{adj} & $-$ & $-$\\
        $(a_{\alpha \dot{\beta}}, \chi^{A}_{\alpha})$ & hyper & \textbf{adj} & $-$ & $-$\\
        $(q_{\dot{\alpha}}, \psi^A)$ & hyper & $\mathbf{k}$ & $\overline{\mathbf{N}}$ & $-$\\
        $(\Xi_l)$ & Fermi & $\mathbf{k}$ & $-$ & $\mathbf{\overline{N}_f}$\\
      \end{tabular}    
    }
    \caption{UV gauge theory on $SU(N)$ self-dual strings in the M-string Higgsing chain}
    \label{fig:2d-m-string-quiver}
\end{figure}

Instanton strings of the 6d $SU(N)$ gauge theory are also self-dual strings of the tensor multiplet. These self-dual strings are realized as $k$ D2-brane segments connecting two NS5-branes, lying along the $016$ directions and on top of the $N$ D6-branes. The separation of two NS5-brane is taken to be very short, so the worldvolume theory of $k$ D2-branes becomes effectively given by the 2d supersymmetric Yang-Mills theory. The D6- and NS5-branes preserve  $SO(5,1)_{012345} \times SO(3)_{789}$ global symmetry, which is further decomposed to $SO(1,1)_{01} \times SO(4)_{2345} \times SO(3)_{789} = SO(1,1) \times SU(2)_l \times SU(2)_r \times SU(2)_R$ due to the presence of D2-branes. Here $SU(2)_R=SO(3)_{789}$ is the R-symmetry of 6d $(1,0)$ SUSY. 
We now examine the preserved supercharges in this system. Let us denote the doublet indices of $(SU(2)_l , SU(2)_r , SU(2)_R)$ symmetry by $(\alpha, \dot{\alpha}, A)$. 32 supercharges of type IIA string theory are decomposed into $Q^{\alpha A}_{\pm\pm}$ and $Q^{\dot{\alpha} A}_{\pm\pm}$, in which the two $\pm$ subscripts denote eigenvalues of $\Gamma^{01}$ and $\Gamma^6$ respectively. Introduction of D2, D6, NS5-branes imposes the SUSY projectors $\Gamma^{016}$, $\Gamma^{0123456}$, $\Gamma^{012345}$, so that 4 supercharges $Q_{-+}^{\dot{\alpha} A}$ are preserved. Since $Q_{-+}^{\dot{\alpha} A}$ have a definite 2d chirality, they form the $\mathcal{N} =(0,4)$ SUSY whose R-symmetry corresponds to $SU(2)_r \times SU(2)_R = SO(4)$. We summarize the field contents of 2d $(0,4)$ gauge theory in Figure~\ref{fig:2d-m-string-quiver}, which are determined from massless modes of open strings connecting D2-D2 and D2-D6. These match with the ADHM construction. In $(0,4)$ quiver diagrams, we draw a solid line to represent a $(0,4)$ hypermultiplet (either twisted or untwisted) and a dashed line to represent a $(0,4)$ Fermi multiplet. 

We make a few comments on the ADHM gauge theory. First, the ADHM gauge theory is originally intended for the $U(N)$ instantons instead of $SU(N)$. It involves extra degrees of freedom, regularizing singular $U(1)$ instantons \cite{Kim:2011mv}. Second, $\mathcal{N}=(0,4)$ chiral SUSYs are preserved so that one should check if the ADHM gauge theory is sensible by looking at a gauge anomaly. In the $(0,4)$ SUSY, fermions are right-moving if they are in hypermultiplets, and left-moving if they are in vector or Fermi multiplets.
Using the normalization $\text{tr}_{\rm fnd}\, (T^a  T^b ) = \frac{1}{2}\delta^{ab}$, the gauge anomaly is proportional to
\begin{align}
     - 4 \times k +  4\times k + 2N \times \tfrac{1}{2}  - N_f \times \tfrac{1}{2}
\end{align}
which vanishes at $N_f = 2N$. We also need to consider a mixed anomaly between gauge and global symmetries. Let us denote the Abelian generators of $U(1)_G \subset U(N)$, $U(1)_F \subset U(N_f)$, and $U(1) \subset U(k)$ by $G$, $F$, and $T_{\rm U(1)}$. Gauge-global mixed anomalies are given as follows.
\begin{align}
  \text{Tr}\, (\gamma_3 \, T_{\rm U(1)}\, F) &= N_f \nonumber\\
  \text{Tr}\, (\gamma_3 \, T_{\rm U(1)}\, G) &= - 2N. \nonumber
\end{align}
These imply that the 2d quantum theory at $N_f=2N$ preserves the $U(1)$ combination $F + G$ of $U(1)_G \times U(1)_F$ only. Third, there exists the fake $U(1)$ symmetry generated by $T_{\rm U(1)} + F + G$, which rotates no fields in the theory.

In the above brane set-up, the Higgs mechanism can be regarded  as a process of decreasing the number of D6-branes. If $N$ reaches to be $1$, the 6d SCFT no longer has a gauge symmetry. The M-theory uplift of the brane configuration is simply two M5-branes supported at the origin of Taub-NUT.  In this case, the ADHM gauge theory turns out to be the $(0,4)$ gauge theory description for M-strings  introduced in \cite{Haghighat:2013fz}. A further Higgsing   removes the last D6-brane, leaving $k$ D2-branes suspended between two NS5-branes. It is also uplifted to M-theory as a pair of parallel M5-branes. The SUSY gauge theory on D2-branes then becomes 2d $(4,4)$ $U(k)$ gauge theory again describing M-strings. In order to study the spectrum of M-string theory, we prefer the $(0,4)$ description to the $(4,4)$ description.
It is because only the first one realizes the $SO(4)$ global symmetry of M-string theory, which rotates 4 transverse directions to M5-branes. Let us decompose it into $SO(4) = SU(2)_L \times SU(2)_R$. At $N=1$, $SU(2)_R$ appears as a part of $(0,4)$ R-symmetry, and also $SU(2)_L$ appears as a $SU(N_f)$ global symmetry. At $N=0$, only $SU(2)_D \subset SU(2)_L \times SU(2)_R$ is realized as a subgroup of $SO(4)$ R-symmetry in $(0,4)$ SUSY.

We now move to the Higgsing sequence \eqref{eq:higgsing-sequence-e-ch2} to the E-string theory. The 6d $SU(3)$ and $SU(2)$ SCFTs in \eqref{eq:higgsing-sequence-e-ch2} have 6 more fundamental hypermultiplets than those 6d SCFTs in \eqref{eq:higgsing-sequence-m-ch2}. As a 6d fundamental hypermultiplet induces a $(0,4)$ Fermi multiplet in the 2d ADHM construction, one may naively attempt to describe $k$ self-dual strings by adding 6 more Fermi multiplets to Figure~\ref{fig:2d-m-string-quiver}. However, this trial immediately fails because a gauge anomaly vanishes only if $N_f = 2N$. Instead, we make the following observations:
\begin{itemize}
  \item $SU(3)$ fundamental hypermultiplets are not distinguishable from $\text{\emph Anti}\,(\bar{\mathbf{3}}\otimes\bar{\mathbf{3}})$ hypermultiplets. They induce the same instanton zero mode. Nevertheless, they supply different degrees of freedom which are supported at the small instanton singularity in the instanton moduli space. 
   
  \item $SU(2)$ antisymmetric hypermultiplets decouple from the rest of perturbative dynamics. However, inclusion of antisymmetric hypermultiplets changes the way of resolving the singularity in the instanton moduli space, by providing extra degrees of freedom supported at the singular locus.

\end{itemize}
Here we take first the empirical approach in order to find a sensible UV completion of the 2d system. Introduction of a single 6d $SU(N)$ antisymmetric hypermultiplet supplies a single set of the following  2d $(0,4)$ supermultiplets \cite{Shadchin:2005mx}:

\begin{center}
  \begin{tabular}{c | c | cccc}
       Field & Type & $U(k)$ & $SU(N)$ & $U(N_f)$ \\
       \hline
       $(\varphi_{A}, \Phi^{\dot{\alpha}})$ & twisted hyper & \textbf{sym} & $-$ & $-$ \\
       $(\Psi_{\alpha})$ & Fermi & \textbf{anti} & $-$ & $-$ \\
       $(\psi)$ & Fermi & $\mathbf{k}$ & $\mathbf{N}$ & $-$ 
   \end{tabular}  
   \end{center}
These additional fields also contribute to a gauge anomaly. If one denotes the number of 6d antisymmetric matter by $N_a$, there would be $N_a$ copies of additional matter contributions to the anomaly. The total amount of 2d $U(k)$ gauge anomaly turns out to be proportional to 
\begin{align}
-4\times k + 4\times k + 2N \times \tfrac{1}{2} - N_f\times \tfrac{1}{2} + 2N_a \times(k+2) -2N_a \times (k-2) -NN_a \times \tfrac{1}{2},
\end{align}
where each contribution is arranged in order that each multiplet has appeared before.
This anomaly cancels if and only if $N_f= (2-N_a)N+8N_a$. The case $N_a=0$ with $N_f=2N$ was already explained. The case $N_a=1$ with $N_f=N+8$ is being discussed from here on.

It is well-known now that there exist the 6d SCFTs with a $SU(N)$ gauge group with $(N+8)$ fundamental hypermultiplets and $1$ antisymmetric hypermultiplet coupled with a tensor multiplet  \cite{Heckman:2015bfa,Bhardwaj:2015xxa}. Our 2d anomaly condition on $k$ self-dual strings of the 6d gauge theories leads to the same consistency condition, boosting the validity of our empirical analysis. 
What is particular about $N = 2,\,3$ compared with $N\ge 4$ is that an antisymmetric representation is either trivial or equivalent to a fundamental representation. As a 6d gauge theory, it is not a priori obvious whether $(N,N_a,N_f)=(3,1,11)$ is better than $(3,0,12)$. Study on 2d self-dual strings implies that the correct description is $(3,1,11)$. Similar consideration applies to $(N,N_a,N_f)=(2,1,10)$. 
This situation resembles the 5d SCFTs which have the $SU(2)$ gauge symmetry. Recall that inclusion of a 5d $SU(2)$ antisymmetric hypermultiplet, which decoupled in the perturbative gauge theory, was crucial for the correct UV completion of instanton quantum mechanics \cite{Hwang:2014uwa}.

\begin{figure}[h!]
    \centering\vspace{0.3cm}
    \subcaptionbox{}{
        \includegraphics[height=2.7cm]{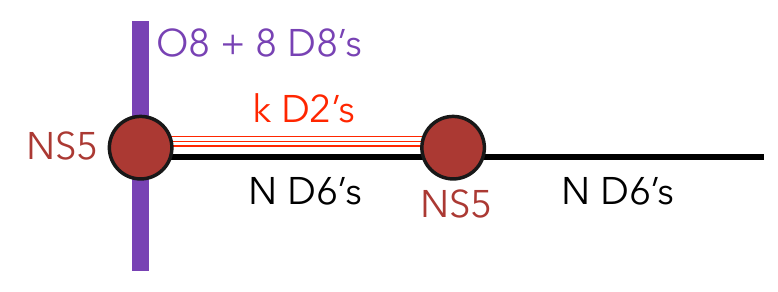}
    }\hfill
    \subcaptionbox{}{
      \begin{tabular}{c|cccccccccc}
        & 0 & 1 & 2 & 3 & 4 & 5 & 6 & 7 & 8 & 9 \\ \hline
        NS5 & $\bullet$ & $\bullet$ & $\bullet$ & $\bullet$ & $\bullet$ & $\bullet$ & $-$ & $-$ &$-$ & $-$\\
        D6 & $\bullet$ & $\bullet$ & $\bullet$ & $\bullet$ & $\bullet$ & $\bullet$ & $\bullet$ & $-$ & $-$ & $-$\\
        O8-D8 & $\bullet$ & $\bullet$ & $\bullet$ & $\bullet$ & $\bullet$ & $\bullet$ & $-$ & $\bullet$ & $\bullet$ & $\bullet$
         \\D2 & $\bullet$ & $\bullet$ & $-$ & $-$ & $-$ & $-$ & $\bullet$ & $-$ & $-$ & $-$
      \end{tabular}  
    }
    \caption{IIA brane system for 6d $SU(N)$ theory Higgsable to E-string theory}
    \label{fig:brane}
\end{figure}

This class $(N,N_a,N_f)=(N,1,N+8)$ of 6d SCFTs can be realized as the IIA brane system \cite{Hanany:1997gh}. Consider the brane set-up in type I$^\prime$ string theory shown in Figure~\ref{fig:brane}. There are a stack of 8 D8-branes on top of an O8$^-$ plane. $N$ semi-infinite D6-branes are extended from the O8$^-$ plane along the $6$ direction. 
There exist a half NS5-brane stuck at the intersection of the O8$^-$ plane and D6-branes, and also a single NS5-brane on D6-branes separated from the O8$^-$ plane. This brane set-up satisfies the requirement of D6-brane charge conservation. 
One can introduce $k$ D2-branes connecting two NS5-branes along the $016$ directions. This brane set-up preserves the symmetry $  SO(1,1) \times SU(2)_l \times SU(2)_r \times SU(2)_R \subset  SO(5,1)_{012345}\times SO(3)_{789} $, which is identical to the symmetry of the brane system in Figure~\ref{fig:m-string-higgsing-sequence}.

If $N$ D6-branes are stretched between two NS5-branes that are very close to each other, the worldvolume dynamics on the D6-brane segments is described by a 6d $SU(N)$ gauge theory with matter multiplets induced from open strings. When open strings connect $N$ D6-brane segments to $N$ semi-infinite D6-branes or D8-branes, they bring fundamental hypermultiplets to the 6d $SU(N)$ gauge theory. When open strings are stretched between $N$ finite D6-branes and their mirror images across the O8$^-$, they induce an antisymmetric hypermultiplet to the 6d $SU(N)$ gauge theory. Adding up, the matter contents of the $SU(N)$ gauge theory are $N_a = 1$ and $N_f = N+8$. 

\begin{figure}[ht]
    \centering
    \subcaptionbox{}{
        \includegraphics[height=4.5cm]{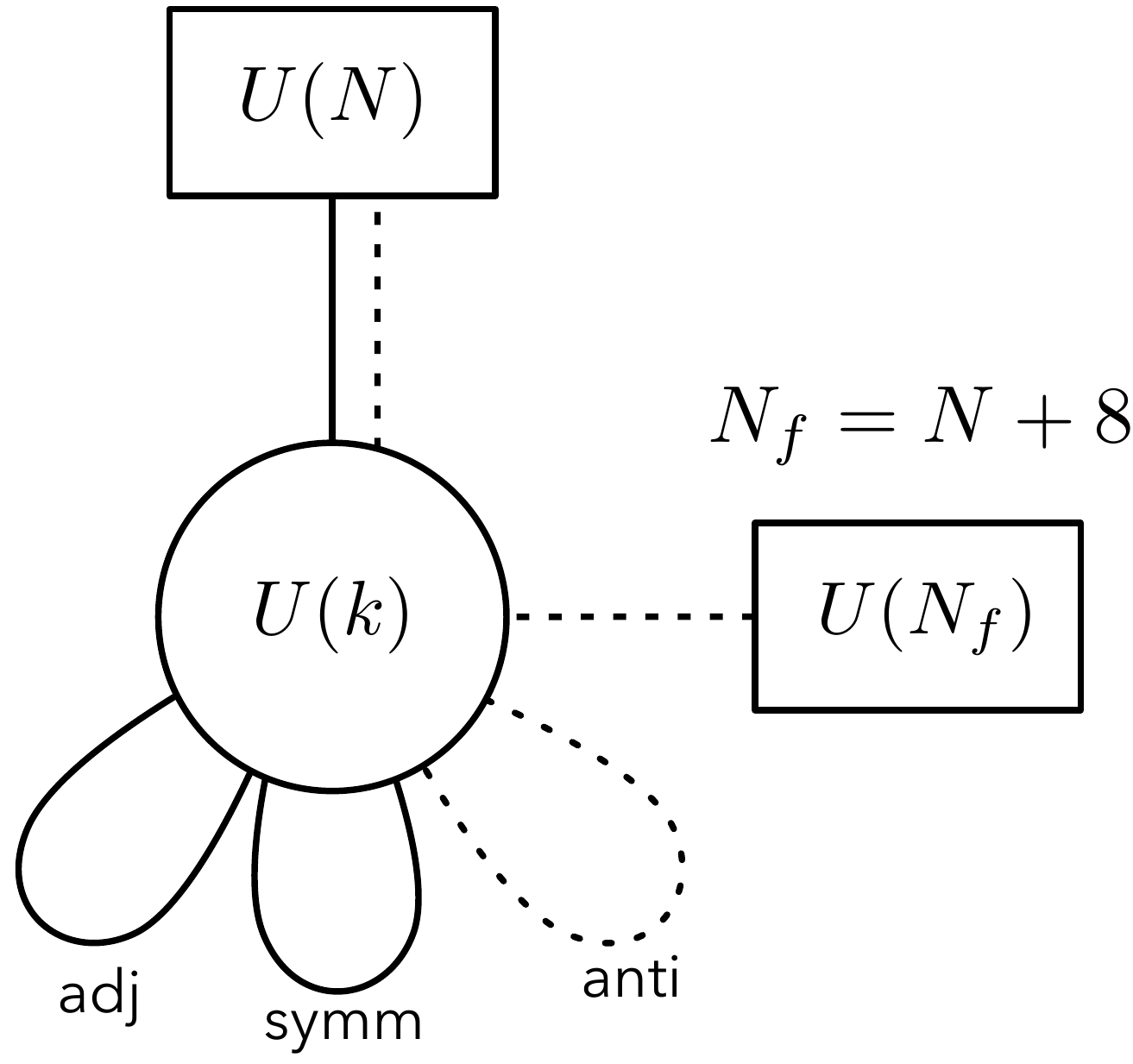}
    }\hfill
    \subcaptionbox{}{
      \begin{tabular}{c| c | cccc}
         Field & Type & $U(k)$ & $U(N)$ & $U(N_f)$ & $U(1)_A$ \\
         \hline
         $(A_\mu, \lambda^{\dot{\alpha}A})$ & vector & \textbf{adj} & $-$ & $-$ & 0\\
         $(a_{\alpha \dot{\beta}}, \chi^{A}_{\alpha})$ & hyper & \textbf{adj} & $-$ & $-$ & 0\\
         $(q_{\dot{\alpha}}, \psi^A)$ & hyper & $\mathbf{k}$ & $\overline{\mathbf{N}}$ & $-$ & 0\\
         $(\Xi_l)$ & Fermi & $\mathbf{k}$ & $-$ & $\mathbf{\overline{N}_f}$ & 0\\
         $(\varphi_{A}, \Phi^{\dot{\alpha}})$ & twisted hyper & \textbf{sym} & $-$ & $-$ & $+1$\\
         $(\Psi_{\alpha})$ & Fermi & \textbf{anti} & $-$ & $-$ & $+1$\\
         $(\psi)$ & Fermi & $\mathbf{k}$ & $\mathbf{N}$ & $-$ & $+1$
      \end{tabular}
    }
    \caption{UV gauge theory on $SU(N)$ self-dual strings in the E-string Higgsing chain}
    \label{fig:2d-quiver}
\end{figure}

Let us explain the 2d gauge theory description for $k$ self-dual strings. Self-dual strings are realized as $k$ D2-branes connecting a half NS5-brane and an NS5-brane, appearing as BPS instantons on top of the D6-brane segments. Among 32 supercharges that are decomposed into $Q^{\alpha A}_{\pm\pm}$ and $Q^{\dot{\alpha} A}_{\pm\pm}$, only 4 supercharges $Q^{\dot{\alpha} A}_{-+}$ remain unbroken because the D2, D6, NS5, D8-branes give rise to the following SUSY projectors: $\Gamma^{016}$, $\Gamma^{0123456}$, $\Gamma^{012345}$, $\Gamma^{012345789}\Gamma^{11} \sim \Gamma^6$. These SUSYs are all right-movers, formulating $\mathcal{N}=(0,4)$ SUSY on the string worldsheet. The worldsheet dynamics of self-dual strings is described by 2d $U(k)$ gauge theory, whose matter contents come from massless modes of open strings connecting $k$ D2-branes to themselves or other neighboring branes. We summarize the field contents in Figure~\ref{fig:2d-quiver}.

Although this 2d gauge theory has no gauge anomaly, one still needs to care about possible mixed anomalies between gauge and global symmetries.
There are three $U(1)$ global symmetries which do not change the Lagrangian explained later in this section. Let us denote the generators of $U(1) \subset U(k)$, $U(1)_F \subset U(N_f)$, $U(1)_A$, $U(1)_G \subset U(N)$ by $T_{\rm U(1)}$, $F$, $A$, and $G$, respectively.  Mixed anomalies are given by
\begin{align}
 \text{Tr}\, (\gamma_3 \, T_{\rm U(1)}\, F) &= N_f \nonumber\\
 \label{eq:mixed-anomaly-ch2}
 \text{Tr}\, (\gamma_3 \, T_{\rm U(1)}\, A) &= 4- N\\
 \text{Tr}\, (\gamma_3 \, T_{\rm U(1)}\, G) &= - 3N, \nonumber
\end{align}
locking an $U(1)$ global symmetry to other two $U(1)$'s in the quantum dynamics. We also comment that there exists the fictitious $U(1)$ symmetry which does not rotate any fields, generated by $T_{\rm U(1)} + G + F - 2A$.

The $N=1$ case is worthy of our attention. This brane system can be uplifted into the heterotic M-theory, as an M5-brane probing an M9-brane intersecting Taub-NUT. Since the corresponding 6d SCFT is the E-string theory, the 2d gauge theory describes the worldsheet dynamics of E-strings. Compared to the $O(k)$ gauge theory description of E-strings \cite{Kim:2014dza}, the above construction is superior in a sense that it realizes the full $SO(4) = SU(2)_L \times SU(2)_R$ global symmetry of E-string theory. While the $O(k)$ gauge theory only sees the diagonal subgroup of $SU(2)_L \times SU(2)_R$, both $SU(2)_R$ and $SU(2)_L$ appear in the $U(k)$ description as a part of $(0,4)$ R-symmetry and the $U(1)_F$ global symmetry. Both descriptions are equally useful when one studies the E-string theory for a single M5-brane near M9-brane, because the BPS spectrum does not include $SU(2)_L$ charged states. However, the $U(k)$ description is more beneficial if one studies E-string theory with many tensor multiplets arising from multiple M5-branes near the M9-brane. Without the $SU(2)_L$ refinement, for example the elliptic genus can count only a fraction of BPS states which factorizes into disjoint E-string theories with 1 tensor multiplet \cite{Gadde:2015tra}.
In Section~\ref{sec:multiple-m5-branes}, we shall study the $SU(2)_L$ flavored spectrum of E-string theory on many M5-branes, refining the result of \cite{Gadde:2015tra}. 

Now we study the Lagrangian of 2d gauge theories in Figure~\ref{fig:2d-quiver}. As there exists no simple formalism which exhibits $(0,4)$ SUSY by construction, to the best of our knowledge, we utilize the $(0,2)$ superfield formalism with the choice of 2 SUSY generators, $Q = Q^{\dot{1}2}_{-+}$ and $Q^\dagger = Q^{\dot{2}1}_{-+}$. We follow the notation of \cite{Benini:2013xpa} in the usage of the $(0,2)$ superfield formalism. After specifying the Lagrangian, we shall content ourselves with observation of the $SO(4)$ R-symmetry, instead of verifying its $(0,4)$ SUSY invariance.
All $(0,4)$ supermultiplets in our gauge theory are decomposed into $(0,2)$ multiplets as follows. 
\begin{align*}
  \text{vector } (A_\mu, \lambda^{\dot{\alpha}A}) &\longrightarrow \text{vector $V$ } (A_\mu, \lambda^{\dot{1}2},\lambda^{\dot{2}1}) + \text{Fermi $\Lambda$ } (\lambda^{\dot{1}1},\lambda^{\dot{2}2})\\
  \text{hyper } (a_{\alpha \dot{\beta}}, \chi^{A}_{\alpha}) &\longrightarrow \text{chiral $B_\alpha$ } (a_{\alpha \dot{1}}, \chi_{\alpha}^{2}) + \text{chiral $\tilde{B}_\alpha^\dagger$ } (a_{\alpha \dot{2}}, \chi_{\alpha}^{1})\\
  \text{hyper } (q_{\dot{\alpha}}, \psi^A) &\longrightarrow \text{chiral $q$ } (q_{\dot{1}}, \psi^{2}) + \text{chiral $\tilde{q}^\dagger$ } (q_{\dot{2}}, \psi^{1})\\
  \text{twisted hyper } (\varphi_{A}, \Phi^{\dot{\alpha}}) &\longrightarrow \text{chiral $\varphi$ } (\varphi_1, \Phi^{\dot{2}}) + \text{chiral $\tilde{\varphi}^\dagger$ } (\varphi_2, \Phi^{\dot{1}})\\
  \text{Fermi $(\Xi_{l}),\,(\Psi_{\alpha}),\, (\psi)$} &\longrightarrow \text{Fermi $(\Xi_{l}),\,(\Psi_{\alpha}),\, (\psi)$ }.
\end{align*}

In the $(0,2)$ formalism, 
the Lagrangian of the system can be completely determined from potentials ($E_\nu$, $J_\nu$) assigned for each $(0,2)$ Fermi multiplet $\nu$. All $E$ and $J$'s are holomorphic functions of $(0,2)$ chiral multiplets, in which the $(0,2)$ SUSY invariance requires
\begin{align}
  \label{eq:02-susy-ej-zero}
  \sum_{\nu \in \text{Fermi}} E_\nu \cdot J_\nu = 0.
\end{align}
The enhanced $(0,4)$ SUSY must be reflected in the way one writes $E$, $J$'s. For example, the $E_{\Lambda}$, $J_{\Lambda}$ functions for the $(0,2)$ Fermi multiplet $\Lambda$ in the $(0,4)$ vector multiplet are constrained as \cite{Tong:2014yna}
\begin{align}
      J_\Lambda &= \sqrt{2}[B_{\alpha}, \tilde{B}^{\alpha}] + \sqrt{2}q\tilde{q} &
  E_\Lambda &= 2 \sqrt{2} \varphi \tilde{\varphi}.
\end{align}
These $E_{\Lambda}$, $J_{\Lambda}$ should be accompanied by $E$, $J$ functions for other Fermi fields to meet the condition \eqref{eq:02-susy-ej-zero}. We turn on the following non-zero $E$, $J$'s for other Fermi fields:
\begin{align}
  J_{\Psi_{1}} &= 2\sqrt{2}[\tilde{\varphi}\tilde{B}^{\alpha} ]_\text{anti} &
  E_{\Psi_{1}} &= 2\sqrt{2}[B_{\alpha} \varphi]_\text{anti} \nonumber\\
  \label{eq:superpotential-without-fi}
  J_{\Psi^{\dagger2}} &= 2\sqrt{2} [\tilde{B}^{\alpha}\varphi]_\text{anti} & 
  E_{\Psi^{\dagger 2} } &= -2\sqrt{2} [\tilde{\varphi}B_{\alpha}]_\text{anti} \\
  J_{\psi} &= 2 (\tilde{\varphi}q)^T & E_{\psi} &= - 2(\tilde{q}\varphi)^T, \nonumber
\end{align}
where $[M]_{\rm anti}$ denotes matrix antisymmetrization, i.e., $[M]_{\rm anti} = \tfrac{1}{2}(M - M^T)$. We find that these $E$, $J$'s satisfy the $(0,2)$ SUSY invariance condition
\begin{align}
  \sum_{\nu \in \text{Fermi}} E_\nu \cdot J_\nu = \text{tr} \left[ -4(\tilde{\varphi}\tilde{B}^{\alpha})(B_{\alpha} \varphi)^T + 4(\tilde{B}^{\alpha} \varphi)(\tilde{\varphi} B_{\alpha})^T \right] = 0
\end{align}
using $\varphi = \varphi^T$. The $D$-term potential is given by
\begin{align}
  D = qq^\dagger - \tilde{q}^\dagger \tilde{q} + [B_{\alpha}, B^{\dagger \alpha}] + [\tilde{B}^{\alpha}, \tilde{B}^{\dagger}{}_{\alpha}] + 2 \varphi \varphi^\dagger - 2\tilde{\varphi}^\dagger \tilde{\varphi}.
\end{align}
Combining all $E$, $J$'s and $D$, one obtains the bosonic potential $V = \tfrac{1}{2}D^2 + \sum_{\nu} (|E_\nu|^2 + |J_\nu|^2) $,
\begin{align}
  \label{eq:potential-without-fi}
  V =&\ \tfrac{1}{2}\left(q_{\dot{\alpha}} (\sigma^{m})^{\dot{\alpha}}{}_{\dot{\beta}} (q^\dagger)^{\dot{\beta}}+ (\sigma^{m})^{\dot{\alpha}}{}_{\dot{\beta}}[a_{\alpha \dot{\alpha}},(a^\dagger)^{\alpha \dot{\beta}}]  \right)^2 +
  2\left(\varphi_{A} (\sigma^{m})^{A}{}_{B} (\varphi^\dagger)^{B}\right)^2 \\
  &\ + \, 2|(q^\dagger)_{\dot{\alpha}} \varphi_A|^2 + \, 2|a_{\alpha\dot{\beta}} \varphi_A - \varphi_A a^T_{\alpha \dot{\beta}}|^2 + \, 2| \varphi_A^\dagger a_{\alpha\dot{\beta}} - a_{\alpha\dot{\beta}}^T \varphi_A^\dagger|^2 \nonumber. 
\end{align}
This is manifestly invariant under an $SO(4)$ R-symmetry transformation of $(0,4)$ SUSY.

One can examine the classical Higgs branch from the bosonic potential $V$. The first term in the potential $V$ imposes the following ADHM constraint
\begin{align}
    q_{\dot{\alpha}} (\sigma^{m})^{\dot{\alpha}}{}_{\dot{\beta}} (q^\dagger)^{\dot{\beta}}+ (\sigma^{m})^{\dot{\alpha}}{}_{\dot{\beta}}[a_{\alpha \dot{\alpha}},(a^\dagger)^{\alpha \dot{\beta}}]  = 0
\end{align}
on the Higgs branch. We claim that the Higgs branch is actually given by an instanton moduli space. If we are away from the small instanton singularity, all $\varphi_A$ fluctuations become massive due to the third term $2|(q^\dagger)_{\dot{\alpha}} \varphi_A|^2$ in the potential $V$. Even at the small instanton singularity, the term proportional to $(|\varphi|^2 + |\tilde{\varphi}|^2)^2$ which comes from the second term in $V$ suppresses a development of the $\varphi_A$ moduli space, although $\varphi$, $\tilde{\varphi}$ are massless there. Therefore, $\varphi_A$ is simply a localized degree of freedom at the small instanton singularity.

We can deform this theory by turning on the FI parameter $\xi^A$ which forms an $SU(2)_R$ triplet. This $\xi^A$ originates from the linear hypermultiplet of 6d $U(N)$ gauge theory, whose VEVs appear as background parameters in the 2d gauge theory.
The 6d theory has the following D-term equations
\begin{align}
  \xi_\mathbb{C} \,\mathbf{1}_N \equiv \frac{\xi^1 + i\xi ^2}{\sqrt{2}}\,\mathbf{1}_N = 2 \sqrt{2} \Phi \tilde{\Phi}, \hspace{0.3cm}
  \xi_\mathbb{R}\,\mathbf{1}_N \equiv \xi^3\, \mathbf{1}_N = 2 \Phi \Phi^\dagger - 2\tilde{\Phi}^\dagger \tilde{\Phi}
\end{align} 
where the 6d antisymmetric hypermultiplet is denoted as the $SU(2)_R$ doublet $(\Phi, \tilde{\Phi}^\dagger)$. They can appear in the 2d gauge theory as background multiplets. In fact, the FI deformation of the 2d gauge theory modifies the holomorphic potentials and D-terms as follows.
\begin{align}
  J_{\psi} = 2(\tilde{\varphi}q)^T + 2\tilde{\Phi}\tilde{q}, \hspace{0.3cm} E_{\psi} = -2(\tilde{q}\varphi)^T + 2q\Phi, \hspace{0.3cm} E_{\Lambda} = 2 \sqrt{2} \varphi \tilde{\varphi} - \xi_\mathbb{C} \mathbf{1}_k, \hspace{0.3cm} D \rightarrow D - \xi_\mathbb{R} \mathbf{1}_k.
\end{align} 
The SUSY invariance condition \eqref{eq:02-susy-ej-zero} still holds, i.e.,
\begin{align}
  \sum_{\nu \in \text{Fermi}} E_\nu \cdot J_\nu = \text{tr} \left[ 4(\tilde{\varphi}q)^T q \Phi - 4(\tilde{q}\varphi)^T \tilde{\Phi}\tilde{q} \right] = 0.
\end{align} 
using $\Phi = -\Phi^T$. The bosonic potential $V$ is also modified as 
\begin{align}
  V =&\ \tfrac{1}{2}\left(q_{\dot{\alpha}} (\sigma^{i})^{\dot{\alpha}}{}_{\dot{\beta}} (q^\dagger)^{\dot{\beta}}+ (\sigma^{i})^{\dot{\alpha}}{}_{\dot{\beta}}[a_{\alpha \dot{\alpha}},(a^\dagger)^{\alpha \dot{\beta}}]  \right)^2 +
  \tfrac{1}{2}\left(2\varphi_{A} (\sigma^{m})^{A}{}_{B} (\varphi^\dagger)^{B} - \xi^m \right)^2 \\
  &\ + 2| (q^\dagger)_{\dot{\alpha}} \varphi_A - \Phi_A (q^T)_{\dot{\alpha}}|^2 + \,2 |a_{\alpha\dot{\beta}} \varphi_A - \varphi_A a^T_{\alpha \dot{\beta}}|^2 + \, 2| \varphi_A^\dagger a_{\alpha\dot{\beta}} - a_{\alpha\dot{\beta}}^T \varphi_A^\dagger|. \nonumber
\end{align}

\begin{figure}[h]
    \centering\vspace{0.3cm}
    \subcaptionbox{$SU(2)$ self-dual string}{
        \includegraphics[width=7cm]{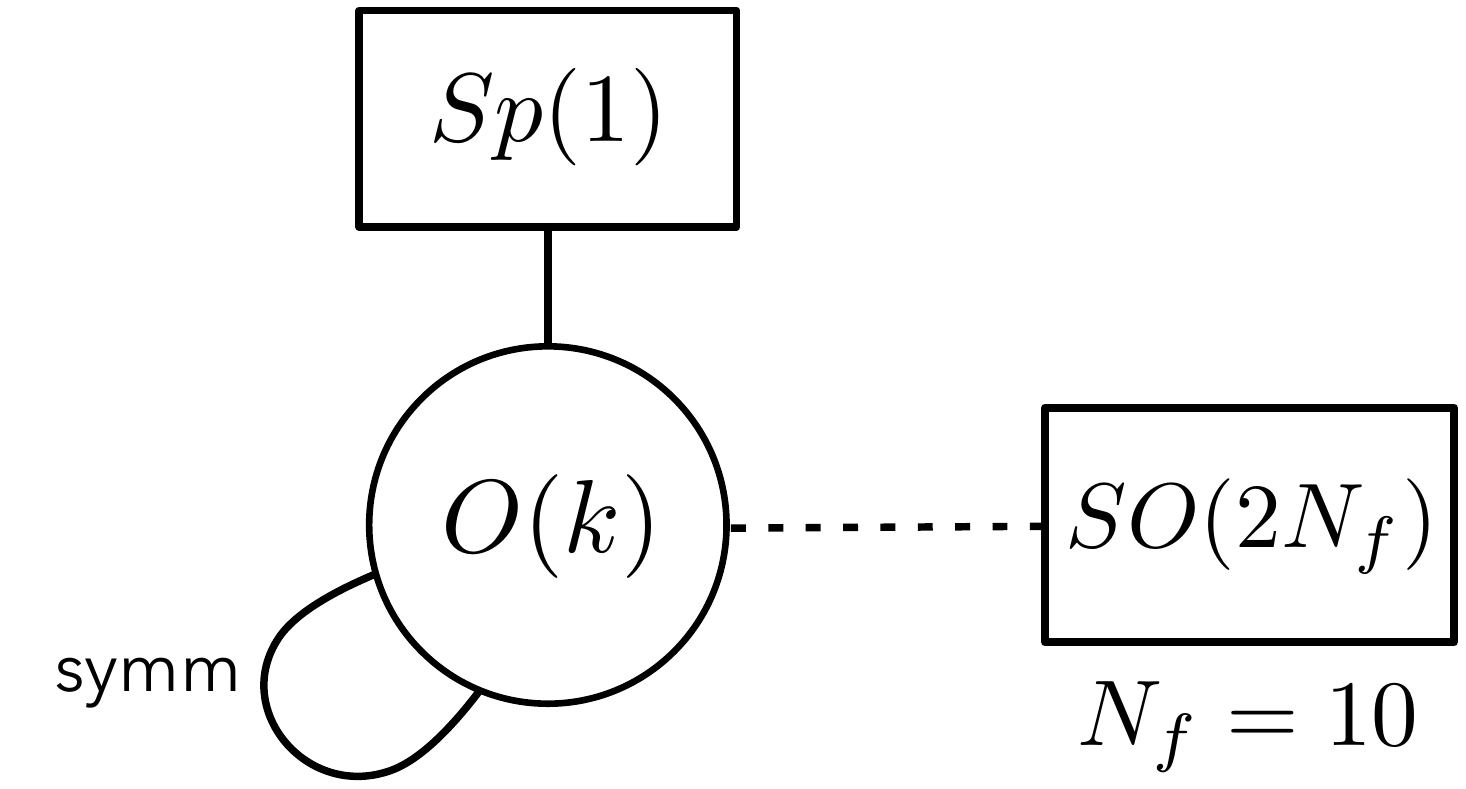}
    }\hfill
    \subcaptionbox{E-string}{
      \includegraphics[width=7cm]{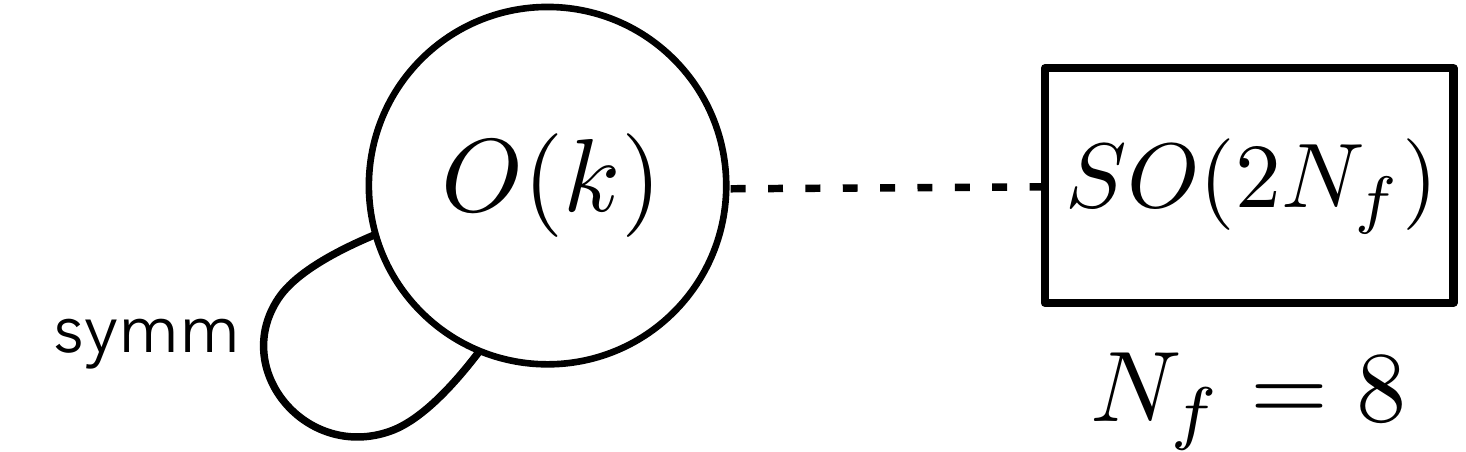}
    }
    \caption{UV gauge theory obtained via the FI deformation of Figure~\ref{fig:2d-quiver}}
    \label{fig:higgsed-ok}
\end{figure}

The Higgs branch may change due to the FI deformation. Let us denote an energy scale by $\mu$ and a 2d gauge coupling by $g$. If the FI deformation is large enough, i.e., $|\xi| \gg \mu \cdot g^{-1}$, some $\varphi_A$ fluctuations become very heavy so that $\varphi_A$ localizes to its classical value, which is proportional to $\sqrt{|\xi|} \,\mathbf{1}_k$. It causes some matrix components of the $U(k)$ gauge field to obtain large masses. We find that $O(k) \subset U(k)$ components of the gauge field remain to be massless, yielding an $O(k)$ gauge theory in the infrared regime. As the FI deformation creates a non-zero VEV of 6d antisymmetric hypermultiplet $(\Phi, \tilde{\Phi}^\dagger)$, it also induces a spontaneous breaking of 6d $SU(N)$ gauge symmetry to $Sp(\frac{N}{2})$ for even $N$ and $Sp(\frac{N-1}{2})$ for odd $N$. The Higgs branch thus becomes an instanton moduli space of 6d $Sp$-type gauge theory. For example, one obtains by the FI deformation the $O(k)$ gauge theories drawn in Figure~\ref{fig:higgsed-ok} which alternatively describe $SU(2)$ self-dual strings and E-strings. 

There also exists another FI deformation which is only compatible with $(0,2)$ SUSY. It is done by
\begin{align}
  D \rightarrow D - \xi_\mathbb{R} \mathbf{1}_k.
\end{align} 
keeping $E$, $J$'s in \eqref{eq:superpotential-without-fi} unchanged. We examine the corresponding Higgs branch for the $k=1$ case. The deformed moduli space is determined by the following set of equations
\begin{align}
    q\tilde{q} = 0, \quad \varphi\tilde{\varphi} = 0, \quad \tilde{\varphi}q = 0, \quad \tilde{q}\varphi = 0, \quad qq^\dagger - \tilde{q}^\dagger \tilde{q} + 2 \varphi \varphi^\dagger - 2\tilde{\varphi}^\dagger \tilde{\varphi} = \xi_{\mathbb{R}}.
\end{align}
Depending on the sign of $\xi_{\mathbb{R}}$, the extra branch of moduli space 
\begin{alignat}{3}
    \tilde{q} = 0, &&\quad \tilde{\varphi} = 0, &\quad qq^\dagger  + 2 \varphi \varphi^\dagger = +\xi_{\mathbb{R}} &&\quad\quad \text{if $\xi_{\mathbb{R}} > 0$}\\
    q = 0, &&\quad \varphi = 0, &\quad \tilde{q}^\dagger\tilde{q} + 2 \tilde{\varphi}^\dagger \varphi = -\xi_{\mathbb{R}} &&\quad\quad \text{if $\xi_{\mathbb{R}} < 0$}
\end{alignat}
can be developed as a novel resolution of the small instanton singularity. It is the weighted projective space $\mathbb{WCP}^N_{1,1,\cdots,1,2}$. This compact branch resolves the degrees of freedom localized at the small instanton singularity, intersecting with the usual instanton moduli space for nonzero $q$ and $\tilde{q}$. 

\section{IR dynamics and symmetry enhancement}
\label{sec:IR-dynamics-symmetry}

It is noticable that the worldsheet gauge theories often do not exhibit the full symmetries of self-dual strings. We especially focus on 6d SCFTs in the Higgsing sequence \eqref{eq:higgsing-sequence-e-ch2} to E-string theory. One expects on self-dual strings some global symmetries, i.e., $SU(12)$ for $G = SU(3)$, $SO(20)$ for $G = SU(2)$, and $E_8$ for E-string theory. However, the UV gauge theory in Figure~\ref{fig:2d-quiver} displays $U(11) \times U(1)_A$, $U(10) \times U(1)_A$, $U(9) \times U(1)_A$ for each case, where $U(1)_F \subset U(N_f)$ and $U(1)_A$ are locked to one another.
 It is thus important to observe the IR global symmetry, which we shall study through the elliptic genera of gauge theories.

Consider the 2d gauge theory on a torus $T^2$ with a complex structure $\tau$. The elliptic genus is defined by picking up $(0,2)$ supercharges, $Q \equiv Q^{\dot{1}2}_-$ and $Q^\dagger \equiv Q^{\dot{2}1}_-$, as follows.
\begin{align}
    \label{eq:elliptic-genus-def}
    Z_k = \text{Tr}_{RR}\,\left[ (-1)^F q^{H_L}\bar{q}^{H_R} e^{2\pi i \epsilon_1 (J_{1} + J_R)} e^{2\pi i \epsilon_2 (J_{2} + J_R)}  e^{2\pi i M A}\prod_{l=1}^{N_f}e^{2\pi i  m_l F_l} \prod_{i=1}^{N}e^{2\pi i  a_i G_i} \right].
\end{align}
Let us explain the Cartan generators and chemical potentials used in \eqref{eq:elliptic-genus-def}. The complex structure $\tau$ appears in \eqref{eq:elliptic-genus-def} through $q \equiv e^{2 \pi i \tau}$. $J_1$ and $J_2$ are Cartans of $SO(4)_{2345}$ rotation, each of which rotates a different 2-plane. $\epsilon_{1,2}$ are chemical potentials for the $SO(4)_{2345}$ angular momentum. $SO(4)_{2345}$ can be decomposed into $SU(2)_l \times SU(2)_r$ whose Cartan generators are denoted as $J_{l}$ and $J_r$. We sometimes replace $\epsilon_{1,2}$ by $\epsilon_{\pm} \equiv \frac{\epsilon_1 \pm \epsilon_2}{2}$ satisfying $\epsilon_1 J_1 + \epsilon_2 J_2 = 2\epsilon_- J_l + 2\epsilon_+ J_r$. $J_R$ is the Cartan generator of $SO(3)_{789}$ rotation which corresponds to $SU(2)_R$ R-symmetry of 6d $\mathcal{N}=(1,0)$ SUSY. For the global symmetry $U(N_f) \times U(1)_A \times U(N)$, we introduce $\{ m_l, M, a_i \}$ and $\{ F_l, A, G_i\}$ to denote associated chemical potentials and Cartan generators. As all generators commute with $Q$, $Q^\dagger$, and $H_R = \{Q, Q^\dagger\}$, the elliptic genus $Z_k$ is independent of $\bar{q}$.

The above chemical potentials are not all independent. One of three global $U(1)$ symmetries is locked to other $U(1)$'s, so that there remain two $U(1)$ combinations which are free of mixed anomalies with a gauge $U(1) \subset U(k)$. Namely, \eqref{eq:mixed-anomaly-ch2} is translated to the following relation among the $U(1)$ chemical potentials:
\begin{align}
  \label{eq:mixed-anomaly-elg}
   (4-N)M + N_f\, \overline{m} - 3 N \, \overline{a} \, =  \, 0,
\end{align}
where $\overline{m}$ and $\overline{a}$ are the $U(1)_F \subset U(N_f)$ and $U(1)_G \subset U(N)$ chemical potentials, defined as
\begin{alignat}{2}
    \label{eq:mixed-anomaly-ch3}
    m_l &= \tilde{m}_l + \overline{m} && \quad \text{ where } \quad \textstyle\sum_{l=1}^{N_f} \tilde{m}_l = 0\\
    a_l &= \tilde{a}_l + \overline{a} && \quad \text{ where } \quad \textstyle\sum_{i=1}^{N} \tilde{a}_i = 0.\nonumber
\end{alignat}
We shall also utilize the fictitious $U(1)$ symmetry to impose $SU(N)$ traceless condition, $\overline{a} = 0$.

The elliptic genus of $\mathcal{N}=(0,2)$ gauge theory was studied in \cite{Benini:2013nda,Benini:2013xpa}. Beginning from the path integral representation, \cite{Benini:2013nda,Benini:2013xpa} developed the general expression of elliptic genera via SUSY localization. We shall apply the result of \cite{Benini:2013nda,Benini:2013xpa} below. One first needs to identify compact zero modes, which comes from the flat connections on $T^2$. By examining all commuting pairs of $U(k)$ group elements, we find that they take the form of
\begin{align}
    A^t = \text{diag}\,(e^{i u_{1}^t},e^{i u_{2}^t},\cdots,e^{i u_{k}^t})\ , \quad
    A^s = \text{diag}\,(e^{i u_{1}^s},e^{i u_{2}^s},\cdots,e^{i u_{k}^s}).
\end{align}
where the superscripts $t$ and $s$ mean temporal and spatial components of the gauge field. The compact zero modes are Wilson lines of the gauge field, written as
\begin{align}
    \phi_i = u_i^{t} + \tau u_i^{s} \quad (i = 1,\cdots,k).
\end{align}
Next we integrate over massive fluctuations, holding zero modes fixed.
It results in a 1-loop determinant factor for each $(0,2)$ supermultiplet. The detailed expression of a 1-loop determinant varies as follows, depending on type and representation $R$ of a given $(0,2)$ multiplet \cite{Benini:2013nda,Benini:2013xpa}.
\begin{align}
    Z_{\rm vector} &=  d \phi_1 \wedge \cdots \wedge d \phi_k \cdot  (2\pi \eta^2)^k\prod_{\alpha \in \text{root}} \frac{\theta_1 (\alpha (\phi))}{i\eta}\\
    Z_{\rm chiral} &= \prod_{\rho \in \text{weight}(R)} \frac{i\eta}{\theta_1 (\rho (\phi) + 2\epsilon_-\, J_l +2\epsilon_+\, (J_r + J_R)  + z\cdot F)}\\
    Z_{\rm Fermi} &= \prod_{\rho \in \text{weight}(R)} \frac{\theta_1 (\rho (\phi) + 2\epsilon_-\, J_l +2\epsilon_+\, (J_r + J_R) + z\cdot F)}{i\eta},
\end{align}
Here we assume the definition of \cite{Kim:2014dza} for the Dedekind eta function $\eta$, the Jacobi theta function $\theta_I (z) \equiv \theta_I (\tau, z)$\ \ ($I = 1, 2, 3, 4$). We collect all 1-loop determinants that correspond to the field contents of our gauge theory.
\begin{align}
    Z_{\rm 1-loop} = Z_{\rm vector} \prod_{\rm chiral} Z_{\rm chiral} \prod_{\rm Fermi} Z_{\rm Fermi}.
\end{align}
Finally, we integrate over zero modes. \cite{Benini:2013nda,Benini:2013xpa} showed that the zero mode integral turns into a contour integral, which needs to be performed along a chosen contour. \cite{Benini:2013nda,Benini:2013xpa} also found that the integral contour is properly determined by the Jeffrey-Kirwan residue,
\begin{align}
  \frac{1}{|W|}\oint Z_{\rm{1-loop}}=
  \frac{1}{|W|}\sum_{\phi_\ast}\textrm{JK-Res}_{\phi_\ast}({\bf Q}_\ast,\mathfrak{n})\,
  Z_{\rm{1-loop}}\ ,
\end{align}
where $\phi_\ast$ runs over all existing poles in the integrand $Z_{\rm 1-loop}$. $W$ is the Weyl subgroup of a gauge group. JK-Res is a linear functional which refers to an auxiliary vector $\mathfrak{n}$ in the $k$-dimensional charge space. ${\bf Q}_\ast=(Q_1,\cdots,Q_r)$ is a set of $r \geq k$ charge vectors, associated to all hyperplanes intersecting at $\phi_\ast$. JK-Res$_{\phi_\ast}({\bf Q}_\ast,\mathfrak{n})$ is defined as
\begin{align}
\label{eq:JK-Res}
  \textrm{JK-Res}_{\phi_\ast}({\bf Q}_\ast,\mathfrak{n})
  \frac{d\phi_1\wedge \cdots\wedge d\phi_r}{Q_{j_1}(\phi\!-\!\phi_\ast)\cdots Q_{j_r}(\phi\!-\!\phi_\ast)}
  =\left\{\begin{array}{ll} | \det(Q_{j_1},\cdots,Q_{j_r}) |^{-1}&
  \textrm{if }\mathfrak{n}\in{\rm Cone}(Q_{j_1},\cdots, Q_{j_r})\\
  0&{\rm otherwise}\end{array}\right..
\end{align}
where $\mathfrak{n}$ has to be generic for JK-Res$_{\phi_\ast}({\bf Q}_\ast,\eta)$ to be well-defined. In this paper, we only encounter non-degenerate poles which arise at intersection points of $k$ hyperplanes. One may consult to \cite{Benini:2013xpa} for treatment of degenerate poles at which $r>k$ hyperplanes intersect.

\paragraph{One string} Following the above instruction, the $k=1$ elliptic genus of the gauge theory in Figure~\ref{fig:2d-quiver} is given by the following contour integral.
\begin{align}
  \label{eq:elliptic-genus-one}
   \oint d\phi\ \frac{\eta^3\, \theta_1(2\epsilon_+)}{i \theta_1(\epsilon_1)\theta_1(\epsilon_2)}
   \cdot \prod_{i=1}^{N} \frac{\eta \, \theta_1 (\phi +a_i + M)}{\theta_1 (\epsilon_+ \pm (\phi-a_i))} 
   \cdot \frac{\eta^2}{\theta_1(-\epsilon_+ \pm (2\phi	+M))} \cdot
   \prod_{l=1}^{N+8} \frac{\theta_1(\phi- m_l)}{\eta}
\end{align}
where we use the notation $\theta_I (a \pm b) \equiv \theta_I (a+b)\,\theta_I (a-b)$ for brevity. The Weyl group $W \subset U(1)$ is trivial. If we pick an auxiliary charge vector $\mathfrak{n}$ to be $+1$, JK-Res corresponds to choose all poles associated with a positive $U(1)$ charge $Q>0$. There are $(N+1)$ relevant poles 
\begin{align}
    \epsilon_+ + \phi - a_j = 0 \quad (j = 1, \cdots, N), \quad\quad -\epsilon_+ + 2 \phi + M = 0,
\end{align}
whose residues are listed below.
\begin{itemize}
	\item $\phi = a_j - \epsilon_+$ \quad ($j = 1, \cdots, N$)
	\begin{align}
	    -\sum_{j=1}^N \frac{\eta^{-6} \, \prod_{l=1}^{N+8} \theta_1 (a_j - \epsilon_+ - m_l) }{\theta_1(\epsilon_1)\theta_1(\epsilon_2) \theta_1 (2 a_j - 3\epsilon_+ + M)}\cdot \prod_{i\neq j} \frac{\theta_1(a_i + a_j -\epsilon_+ + M)}{\theta_1(a_j-a_i) \theta_1(2\epsilon_+ - (a_j - a_i))} 
	\end{align}
	\item $\phi = \frac{\epsilon_+ - M}{2} + \ell_I$ for $\ell = \{0, \frac{1}{2},\frac{1+\tau}{2},\frac{\tau}{2}\}$ \quad ($I=1,2,3,4$)
	\begin{align}
	    -\frac{1}{2}  \frac{\eta^{-6}}{\theta_1 (\epsilon_1) \theta_1 (\epsilon_2)}  \Bigg[\frac{\prod_{l=1}^{N+8} \theta_1 (\frac{\epsilon_+ - M}{2} - m_l)}{ \prod_{i=1}^{N} \theta_1 ( \frac{3\epsilon_+ - M}{2} - a_i)} + (-1)^N \sum_{I=2}^4 \frac{\prod_{l=1}^{N+8} \theta_I (\frac{\epsilon_+ - M}{2} - m_l)}{ \prod_{i=1}^{N} \theta_I ( \frac{3\epsilon_+ - M}{2} - a_i)} \Bigg]
	\end{align}
\end{itemize}

\paragraph{Two strings} The $k=2$ elliptic genus of the gauge theory in Figure~\ref{fig:2d-quiver} can be expressed as
\begin{align}
  \label{eq:elliptic-genus-two}
   \oint \frac{d\phi_{1,2}}{2}\ &\frac{- \eta^6 \theta_1 (2 \epsilon_+)^2}{\theta_1(\epsilon_1)^2 \theta_1 (\epsilon_2)^2}\prod_{i\neq j} \frac{\theta_1(\phi_{ij})\theta_1(\phi_{ij} + 2 \epsilon_+)}{\theta_1( \phi_{ij} + \epsilon_1) \theta_1(\phi_{ij} + \epsilon_2)} \prod_{l=1}^{N+8} \frac{\theta_1 (\phi_{1,2}-m_l)}{\eta^2} \prod_{i=1}^N \frac{\eta^2 \theta_1 (\phi_{1,2} + a_i + M)}{\theta_1 (\epsilon_+ \pm (\phi_{1,2}-a_i))}\\
   &\times \frac{\eta^4 \theta_1 (\epsilon_- \pm (\phi_1 + \phi_2+ M))}{\theta_1 (-\epsilon_+ \pm (\phi_1 + \phi_2 + M))\theta_1 (-\epsilon_+ \pm (2\phi_{1,2} + M))}. \nonumber
\end{align}
We adopt the concise notations such as $\phi_{ij} \equiv \phi_i - \phi_j$, $a_{mn} \equiv a_m - a_n$, $\theta_I (\phi_{i,j} + b) \equiv \theta_I (\phi_{i} + b)\, \theta_I (\phi_{j} + b)$, $\theta_I (a_{m,n} + b) \equiv \theta_I (a_m + b)\, \theta_I (a_n + b)$, $\theta_{I,J}(b) \equiv \theta_{I}(b)\, \theta_J (b)$. The Weyl group $W \subset U(2)$ is $\mathbb{Z}_2$. After picking an auxiliary vector $\mathfrak{n}$ to be $(+1, +1)$, we collect all contributing residues given as follows.
\begin{itemize}
  \item $(\phi_1, \phi_2) = (a_m - \epsilon_+, a_n - \epsilon_+)$ for $m\neq n$.
  \begin{align}
      &\sum_{m\neq n}\frac{\eta^4}{2\theta_1(\epsilon_1)^2 \theta_1 (\epsilon_2)^2}\frac{\eta^4}{\theta_1( a_{mn} \pm \epsilon_1) \theta_1(a_{mn} \pm \epsilon_2)} \prod_{l=1}^{N+8} \frac{\theta_1 (a_{m,n}-\epsilon_+-m_l)}{\eta^2} \\
      &\times  \frac{\theta_1 (a_m + a_n - 2 \epsilon_+ \pm \epsilon_- + M) \, \theta_1 ( a_m + a_n - \epsilon_+ + M)}{\theta_1(a_m + a_n - 3 \epsilon_+ + M)\, \theta_1(2a_{m,n} - 3 \epsilon_+ + M)} \prod_{i\neq \{m,n \} }^N \frac{\eta^2 \theta_1 (a_{m,n} + a_i - \epsilon_+ + M)}{\theta_1 (a_{m,n}-a_i) \theta_1 (2\epsilon_+ - (a_{m,n}-a_i))} \nonumber
  \end{align}
  \item $(\phi_1, \phi_2) = (\frac{\epsilon_+ - M}{2} + \ell_I, a_m - \epsilon_+)$ and 
  $(\phi_1, \phi_2) = (a_m - \epsilon_+, \frac{\epsilon_+ - M}{2} + \ell_I)$
  \begin{align}
      &\sum_{I=1}^4 \sum_{m=1}^N \frac{\eta^4}{2\theta_1(\epsilon_{1,2})^2 } \frac{\eta^2 \,\theta_I(\frac{M-7 \epsilon_+}{2} + a_m) }{\theta_1 (M-3 \epsilon_+ + 2a_m) \theta_I (\frac{M-5 \epsilon_+}{2} \pm \epsilon_- + a_m)} \\
      &\times \prod_{l=1}^{N+8} \frac{\theta_1 (-a_m + \epsilon_+ + m_l) \theta_I ( \frac{\epsilon_+ - M}{2} - m_l)}{\eta^2}
       \prod_{i\neq m}^N \frac{ \eta^2 \,\theta_1 (M- \epsilon_+ + a_m + a_i)}{\theta_1 (a_m - a_i) \theta_1 (2 \epsilon_+ - (a_m - a_i)) \theta_I ( \frac{M-3 \epsilon_+}{2} + a_i)} \nonumber
  \end{align}
  \item $(\phi_1, \phi_2) = (\frac{\epsilon_+ - M}{2} + \ell_I, \frac{\epsilon_+ - M}{2} + \ell_J)$
  \begin{align}
      \sum_{(I,J,K)\in \mathcal{S}} \frac{\eta^4\, \theta_K(0) \theta_K (2 \epsilon_+)}{4\theta_1(\epsilon_{1,2})^2\theta_K (\epsilon_{1,2})} \prod_{i=1}^N \frac{\eta^2}{\theta_{I,J} (\frac{M- 3 \epsilon_+}{2}+a_i)}\prod_{l=1}^{N+8}\frac{\theta_{I,J}(\frac{\epsilon_+-M}{2}-m_l)}{\eta^2}
  \end{align}
  where the summation is taken over the set $\mathcal{S} = \{ (1,2,2),(1,3,3),(1,4,4),(2,3,4),(2,4,3),(3,4,2)  \}$.
  \item $(\phi_1, \phi_2) = (a_m - \epsilon_+, a_m - \epsilon_+ - \epsilon_{1,2})$ and $(\phi_1, \phi_2) = (a_m - \epsilon_+ - \epsilon_{1,2}, a_m - \epsilon_+)$
  \begin{align}
      &\sum_m \frac{-\eta^4}{\theta_1(\epsilon_{1,2}) \theta_1(2 \epsilon_1) \theta_1(2 \epsilon_-)} \frac{\eta^2}{\theta_1(2a_m - 3 \epsilon_+ -\epsilon_1 +  M) \, \theta_1(2a_m - 3\epsilon_+ - 2 \epsilon_1 + M )}\\
      &\times\prod_{l=1}^{N+8} \frac{\theta_1 (a_{m}-\epsilon_+-m_l) \theta_1 (a_{m}-\epsilon_+- \epsilon_{1}-m_l)}{\eta^2} \nonumber \\
      &\times\prod_{i\neq m }^N \frac{\eta^2 \theta_1 (a_{m} + a_i - \epsilon_+ + M) \theta_1 (a_{m} + a_i -\epsilon_1 - \epsilon_+ + M)}{\theta_1 (a_{m}-a_i) \theta_1 (2\epsilon_+ - (a_{m}-a_i))\theta_1 (a_{m}-a_i -\epsilon_1) \theta_1 (2\epsilon_+ + \epsilon_1 - (a_{m}-a_i))} + (\epsilon_1 \leftrightarrow \epsilon_2) \nonumber
  \end{align}
  \item $(\phi_1, \phi_2) = (-a_m + 2\epsilon_+ - M, a_m - \epsilon_+)$
  \begin{align}
      &\sum_m \frac{-\eta\,  \theta_1 (2 \epsilon_+)}{2\theta_1(\epsilon_{1,2}) } 
      \frac{\eta^5}{\theta_1(2a_m -3 \epsilon_+  -\epsilon_{1,2} + M) \theta_1(2a_m - \epsilon_+  -\epsilon_{1,2} + M) \theta_1 ( 2a_1 -3 \epsilon_+ + M)}\\
      &\times\prod_{l=1}^{N+8} \frac{\theta_1 (a_{m}-\epsilon_+-m_l) \theta_1 (-a_{m}+2 \epsilon_+ -M -m_l)}{\eta^2} \prod_{i\neq m }^N \frac{\eta^2}{\theta_1 (a_m-a_i) \theta_1 (3 \epsilon_+  - (a_m+ a_i + M))} \nonumber
  \end{align}
  \item 
  $(\phi_1, \phi_2) = (\frac{\epsilon_+ - M}{2} + \frac{\epsilon_{1,2}}{2} + \ell_I , \frac{\epsilon_+ - M}{2} - \frac{\epsilon_{1,2}}{2} + \ell_I )$
  \begin{align}
      \sum_{I=1}^4 &\frac{-\eta^4}{4\theta_1(2 \epsilon_-) \theta_1 (\epsilon_{1,2}) \theta_1(2 \epsilon_1)} \prod_{i=1}^N \frac{\eta^2}{\theta_I (\frac{M- \epsilon_-}{2}-2 \epsilon_+ + a_i) \, \theta_I (\frac{M+ \epsilon_-}{2}- \epsilon_+ + a_i)} \\
      &\times \prod_{l=1}^{N+8}\frac{\theta_I (\frac{M+ \epsilon_-}{2} + m_l) \theta_I (\frac{M- \epsilon_-}{2} - \epsilon_+ + m_l)}{\eta^2} + (\epsilon_1 \leftrightarrow \epsilon_2), \nonumber
  \end{align}
\end{itemize}

\subsection{$SU(3)$ instanton strings with 12 flavors}

Let us examine the elliptic genera of the UV gauge theory in Figure~\ref{fig:2d-quiver} at $N=3$. We first consider the special limit where $\epsilon_+ = 0$ and $a_3 = 0$. Utilizing the fictitious $U(1)$ symmetry to impose the $SU(3)$ traceless condition, $a_3 = 0$ implies $a_1 = -a_2 = a$. We take a series expansion of the $k=1$ elliptic genus in $a$, obtaining
\begin{align}
    &a^{-4} \cdot  \frac{\eta^2}{\theta_1 (\epsilon_-)^2} \frac{3\theta_1 (M) \theta_1(m_1)\cdots \theta_1 (m_{11})}{2 (2\pi)^4\, \eta^{12}} + a^{-2} \cdot \bigg( \sum_{\sigma \in P_1}\frac{\eta^2}{\theta_1 (\epsilon_-)^2} \frac{\theta_1^{(\sigma_1)}(m_1) \cdots \theta_1^{(\sigma_{11})} (m_{11}) \theta_1^{(\sigma_{12})} (M)} {2(2\pi)^4\, \eta^{12}} \nonumber\\&
    +\sum_{\sigma \in P_2}\frac{\eta^2}{\theta_1 (\epsilon_-)^2} \frac{\theta_1^{(\sigma_1)}(m_1) \cdots \theta_1^{(\sigma_{11})} (m_{11}) \theta_1^{(\sigma_{12})} (M)} {4(2\pi)^4\, \eta^{12}} +\frac{\eta^2}{\theta_1 (\epsilon_-)^2} \frac{3 E_2 \,\theta_1 (M) \theta_1(m_1)\cdots \theta_1 (m_{11})}{8 (2\pi)^2 \,\eta^{12}}\bigg) + \mathcal{O}(a^0)
\end{align}
where $\theta_I^{(n)} (z) \equiv (\partial_z)^n \theta_I (z)$. $P_1$ and $P_2$ are sets of all possible permutations of $\{0,0,0,0,0,0,0,0,0,0,1,1\}$ and $\{0,0,0,0,0,0,0,0,0,0,0,2\}$ respectively. $\sigma_i$ denotes an $i$'th element of $\sigma$. Together with \eqref{eq:mixed-anomaly-elg}, $M + \sum_{l=1}^{11} m_l = 0$, we conclude that the above expression displays the expected $SU(12)$ symmetry because it is manifestly invariant under the $SU(12)$ Weyl symmetry which permutes $m_1, \cdots, m_{11}$ and $m_{12} \equiv M$.

Now we inspect the series expansion of the elliptic genera by $q \equiv e^{2\pi i \tau}$ up to a certain order, keeping all the fugacities nonzero. Resulting expressions are written in terms of fugacity variables, which are defined as
\begin{align}
  \label{eq:fugacity-definition}
  t = e^{2 \pi i \epsilon_+},\ u = e^{ 2 \pi i \epsilon_-},\ y_i = e^{ 2 \pi i \tilde{m}_i},\ \overline{y} = e^{2\pi i \overline{m}}, \ Y = e^{ 2 \pi i M}, \ w_i = e^{2\pi i \tilde{a}_i}, \ \overline{w} = e^{2\pi i \overline{a}}.
\end{align}
We take advantage of the fake $U(1)$ symmetry to set $\bar{w} = 0$. The equation \eqref{eq:mixed-anomaly-elg} then relates $Y$ to $\overline{y}$  such that $Y = \bar{y}^{-11}$. Looking at the BPS spectrum after $q$-expansion, we find that the elliptic genera exhibit the global symmetry enhancement from $SU(11) \times U(1)_F$ to $SU(12)$. The branching rules, e.g.,
\begin{align*}
  \mathbf{12} &\longrightarrow \mathbf{1}_{-11}+\mathbf{11}_{+1}\\
  \overline{\mathbf{12}} &\longrightarrow \mathbf{1}_{+11}+\overline{\mathbf{11}}_{-1}\\
  \mathbf{143} &\longrightarrow \mathbf{1}_0 + \mathbf{11}_{12}+\overline{\mathbf{11}}_{-12}+\mathbf{120}_0,
\end{align*}
are used to organize BPS states into $SU(12)$ irreducible representations. For example, the elliptic genus for $k=1$ captures the following spectrum
\begin{align}
    \label{eq:expand-spectrum}
   \frac{-t}{(1-tu)(1-tu^{-1})} &\bigg[q^{-1/2} 
   + q^{1/2}\cdot \Big( t^{-2} \chi_{\mathbf{8}}^{SU(3)} + t^{-1} \chi_{1/2}^{\text{SU}(2)}(u)   - t^{-1}\big(\chi_{\mathbf{3}}^{SU(3)}  \chi^{\rm SU(12)}_{\overline{\mathbf{12}}} + \chi_{\bar{\mathbf{3}}}^{SU(3)}   \chi^{\rm SU(12)}_{\mathbf{12}} )  \nonumber \\
   & + \chi^{\rm SU(12)}_{\mathbf{143}} +1 + \chi_{\mathbf{8}}^{SU(3)}  + \mathcal{O}(t^1)\Big) + \mathcal{O}(q^{3/2})   \bigg].
\end{align}
which displays the $SU(12)$ global symmetry. $\chi^{\rm G}_{\mathbf{R}}$ denotes the character of a representation $R$ in a symmetry group $G$. We also use the symbol $\chi_{j}^{\text{SU}(2)}$ to denote a spin-$j$ representation of $SU(2)$. Note that the overall factor $\frac{t}{(1-tu)(1-tu^{-1})}$ reveals a tower of infinitely many states, carrying different $SO(4)_{2345}$ quantum numbers. It reflects the center-of-mass motion of a self-dual string along the transverse $\mathbb{R}^4$. 
The BPS spectrum \eqref{eq:expand-spectrum} also shows that a zero-point energy of the left-moving Hamiltonian $H_L$ is $-\frac{1}{2}$. One again observes the $SU(12)$ global symmetry at $k=2$,
\begin{align}
   \frac{t^2}{(1-tu)^2(1-tu^{-1})^2} &\bigg[  q^{-1} \cdot \frac{t \,\chi _{1/2}^{\text{SU}(2)}(t) }{(1+ tu^{-1}) (1+ t u)} + q^{0} \cdot \Big(
   t^{-2} \chi_{\mathbf{8}}^{SU(3)} + t^{-1} \big( \chi_{1/2}^{\text{SU}(2)}(u) - \chi_{\mathbf{3}}^{SU(3)}  \chi^{\rm SU(12)}_{\overline{\mathbf{12}}} \nonumber\\
   &+ \chi_{\bar{\mathbf{3}}}^{SU(3)}   \chi^{\rm SU(12)}_{\mathbf{12}}\big) + \chi^{\rm SU(12)}_{\mathbf{143}} +1 + \chi_{\mathbf{8}}^{SU(3)} + \mathcal{O}(t^1) \Big) +  \mathcal{O}(q^1) \bigg].
\end{align}
The $SU(12)$ symmetry enhancement suggests that the 2d gauge theory in Figure~\ref{fig:2d-quiver} properly describe self-dual strings of 6d $SU(3)$ SCFT with 12 flavors in the infrared regime. 

\subsection{$SU(2)$ instanton strings with 10 flavors}

Let us consider the elliptic genera for 6d $SU(2)$ self-dual strings. The fictitious $U(1)$ symmetry allows us to impose the $SU(2)$ traceless condition $\bar{a} = 0$, implying $a_1 = -a_2 = a$. The equation \eqref{eq:mixed-anomaly-elg} then links $Y$ to $\overline{y}$ such that $Y = \bar{y}^{-5}$.

Recall that the FI deformation gives another description of $SU(2)$ strings, which is the $O(k)$ gauge theory in Figure~\ref{fig:higgsed-ok}(a). It has the $Sp(1) \times SO(20)$ global symmetry from the beginning. Listing down all $(0,4)$ matter contents,
\begin{center}
      \begin{tabular}{c| c | ccc}
         Field & Type & $O(k)$ & $Sp(1)$ & $SO(20)$\\
         \hline
         $(A_\mu, \lambda^{\dot{\alpha}A})$ & vector & \textbf{adj} & $-$ & $-$\\
         $(a_{\alpha \dot{\beta}}, \lambda^{A}_{\alpha})$ & hyper & \textbf{sym} & $-$ & $-$\\
         $(q_{\dot{\alpha}}, \psi^A)$ & hyper & $\mathbf{k}$ & $\mathbf{2}$ & $-$ \\
         $(\Xi_l)$ & Fermi & $\mathbf{k}$ & $-$ & $\mathbf{20}$\\
      \end{tabular}
\end{center}
one can check that the gauge anomaly is absent: $-4(2k-2) + 4(2k+2) + 4 - 20 = 0$.
We expect the elliptic genera of both gauge theories should be the same, even if $U(k)$ and $O(k)$ theories are seemingly different. It is because the elliptic genus is insensitive to the FI deformation. Based on the correspondence between $U(k)$ and $O(k)$ elliptic genera, we expect the $SO(20)$ symmetry enhancement to hold. As a sample computation, the elliptic genus of $O(1)$ gauge theory can be written as
\begin{align}
    -\frac{\eta^2}{\theta_1 (\epsilon_{1,2})} \sum_{I=1}^4 \frac{\eta^2}{\theta_I (\epsilon_+ \pm a)} \prod_{l=1}^{10}\frac{\theta_I (m_l)}{\eta}
\end{align}
where we tested the agreement with the $U(1)$ elliptic genus
\begin{align}
    - \frac{\eta^{-6}  }{\theta_1(\epsilon_{1,2})}\Bigg[&\frac{\prod_{l=1}^{10} \theta_1 (a- \epsilon_+ - m_l)}{ \theta_1 (2 a - 3\epsilon_+ + M)} \frac{\theta_1(-\epsilon_+ + M) }{\theta_1 (2a ) \theta_1(2 \epsilon_+ - 2a)}  + (\pm a \rightarrow \mp a)\Bigg] 
      - \frac{\eta^{-6}  }{\theta_1(\epsilon_{1,2})}\sum_{I=1}^4 \frac{\prod_{l=1}^{10} \theta_I (\frac{\epsilon_+ - M}{2} - m_l)}{  2\theta_I ( \frac{3\epsilon_+ - M}{2} \pm a)}
\end{align}
by the series expansion up to $q^{3/2}$ order and also at the point $m_{l} = M = \epsilon_+ = 0$ in all orders of $q$. Similarly, we checked that the $O(2)$ elliptic genus agrees with the $U(2)$ elliptic genus \eqref{eq:elliptic-genus-two} by the series expansion up to $q^2$ order.
This suggests that that the $U(k)$ gauge theory in Figure~\ref{fig:2d-quiver} at $N=2$ is a correct UV description of self-dual strings in 6d $SU(2)$ SCFT with $N_f = 10$.

Taking advantage of the $SO(20) \rightarrow SU(10) \times U(1)_F$ branching rules, e.g.,
\begin{align*}
  \mathbf{20} &\longrightarrow \mathbf{10}_{+1}+\overline{\mathbf{10}}_{-1}\\
  \mathbf{190} &\longrightarrow \mathbf{1}_0 + \mathbf{45}_{+2}+\overline{\mathbf{45}}_{-2}+\mathbf{99}_0\\
  \mathbf{512} &\longrightarrow 
  \mathbf{1}_{+5}+\mathbf{1}_{-5}+\mathbf{45}_{-3}+\overline{\mathbf{45}}_{+3}+\mathbf{210}_{-1}+\overline{\mathbf{210}}_{+1}\\
  \overline{\mathbf{512}} &\longrightarrow \mathbf{10}_{-4}+\overline{\mathbf{10}}_{+4} + \mathbf{120}_{-2} +\overline{\mathbf{120}}_{+2} +\mathbf{252}_{0}.
\end{align*}
we organize the BPS spectum of self-dual strings into $SO(20)$ irreducible representations. At $k=1$,
\begin{align}
   &\frac{t}{(1-tu)(1-tu^{-1})} \bigg[q^{-1/2} 
   + \frac{q^{1/2}\cdot t^2}{(1-t^2 w_1^2)(1-t^2 w_1^{-2})} \Big(
   -\chi ^{\text{SO}(20)}_{\bf\overline{512}} \chi ^{\text{SU}(2)}_{1/2}\left(w_1\right)+\chi ^{\text{SO}(20)}_{\bf 512} \chi
   ^{\text{SU}(2)}_{1/2}(t)\\
   &+\chi ^{\text{SO}(20)}_{\bf 20} \chi
   ^{\text{SU}(2)}_{1/2}(t) \chi ^{\text{SU}(2)}_{3/2}\left(w_1\right)-\chi ^{\text{SO}(20)}_{\bf 20} \chi^{\text{SU}(2)}_{3/2}(t) \chi ^{\text{SU}(2)}_{1/2}\left(w_1\right)-\chi ^{\text{SO}(20)}_{\bf 190} \chi
   ^{\text{SU}(2)}_1\left(w_1\right)+\chi ^{\text{SU}(2)}_{1/2}(t) \chi ^{\text{SU}(2)}_{1/2}(u) \nonumber\\
   &+\chi^{\text{SU}(2)}_{3/2}(t) \chi ^{\text{SU}(2)}_{1/2}(u)-\chi ^{\text{SU}(2)}_{1/2}(t) \chi
   ^{\text{SU}(2)}_{1/2}(u) \chi ^{\text{SU}(2)}_1\left(w_1\right)+\chi ^{\text{SU}(2)}_2(t) \chi
   ^{\text{SU}(2)}_1\left(w_1\right)-\chi ^{\text{SU}(2)}_1(t) \chi ^{\text{SU}(2)}_2\left(w_1\right)\nonumber\\
   & +\chi ^{\text{SO}(20)}_{\bf 190} \chi ^{\text{SU}(2)}_1(t) \Big) + \mathcal{O}(q^{3/2})\bigg]. \nonumber
\end{align}
Similarly at $k=2$, we obtain the following spectrum
\begin{align}
   &\frac{t^2}{(1-tu)^2(1-tu^{-1})^2} \bigg[   \frac{q^{-1} \cdot t \,\chi _{1/2}^{\text{SU}(2)}(t) }{(1+ tu^{-1}) (1+ t u)} + \frac{q^{0}\cdot t^2}{(1-t^2 w_1^2)(1-t^2 w_1^{-2})}\Big(
   -\chi ^{\text{SO}(20)}_{\bf\overline{512}} \chi ^{\text{SU}(2)}_{1/2}\left(w_1\right)+\chi ^{\text{SO}(20)}_{\bf 512} \chi
   ^{\text{SU}(2)}_{1/2}(t) \nonumber\\
   &+\chi ^{\text{SO}(20)}_{\bf 20} \chi
   ^{\text{SU}(2)}_{1/2}(t) \chi ^{\text{SU}(2)}_{3/2}\left(w_1\right)-\chi ^{\text{SO}(20)}_{\bf 20} \chi^{\text{SU}(2)}_{3/2}(t) \chi ^{\text{SU}(2)}_{1/2}\left(w_1\right)-\chi ^{\text{SO}(20)}_{\bf 190} \chi
   ^{\text{SU}(2)}_1\left(w_1\right)+\chi ^{\text{SU}(2)}_{1/2}(t) \chi ^{\text{SU}(2)}_{1/2}(u) \nonumber\\
   &+\chi^{\text{SU}(2)}_{3/2}(t) \chi ^{\text{SU}(2)}_{1/2}(u)-\chi ^{\text{SU}(2)}_{1/2}(t) \chi
   ^{\text{SU}(2)}_{1/2}(u) \chi ^{\text{SU}(2)}_1\left(w_1\right)+\chi ^{\text{SU}(2)}_2(t) \chi
   ^{\text{SU}(2)}_1\left(w_1\right)-\chi ^{\text{SU}(2)}_1(t) \chi ^{\text{SU}(2)}_2\left(w_1\right)\nonumber\\
   & +\chi ^{\text{SO}(20)}_{\bf 190} \chi ^{\text{SU}(2)}_1(t) \Big) + \mathcal{O}(q^1)\bigg].
\end{align}

\section{Muliple M5-branes probing an M9-brane}
\label{sec:multiple-m5-branes}

In this section, we continue our study on the gauge theory elliptic genus. We verify that the 2d gauge theory in Figure~\ref{fig:2d-quiver} can describe E-strings, by checking the $E_8$ symmetry enhancement in the elliptic genera. More importantly, we present quiver gauge theories describing chains of strings, which exist on multiple M5-branes probing an M9-brane. This enables us to study the fully flavored BPS spectrum of E-string theory with a multi-dimensional tensor branch, refining the result of \cite{Gadde:2015tra}.

E-string theory is a 6d SCFT on M5-branes near an M9-brane. When M5-branes are near an M9-brane, M2-branes can be suspended between M9- and M5-branes, or between a pair of M5-branes. These M2-branes induce self-dual strings in E-string theory. If we focus on M2-branes between M9- and M5-branes, the induced strings enjoy the $E_8$ symmetry of the M9-brane, being called E-strings \cite{Ganor:1996mu,Seiberg:1996vs,Klemm:1996dt,Minahan:1998vr}. More generally, there arises a chain of self-dual strings which is induced from a generic configuration of suspended M2-branes. 

\begin{figure}[t]
  \centering
  \includegraphics[height=3cm]{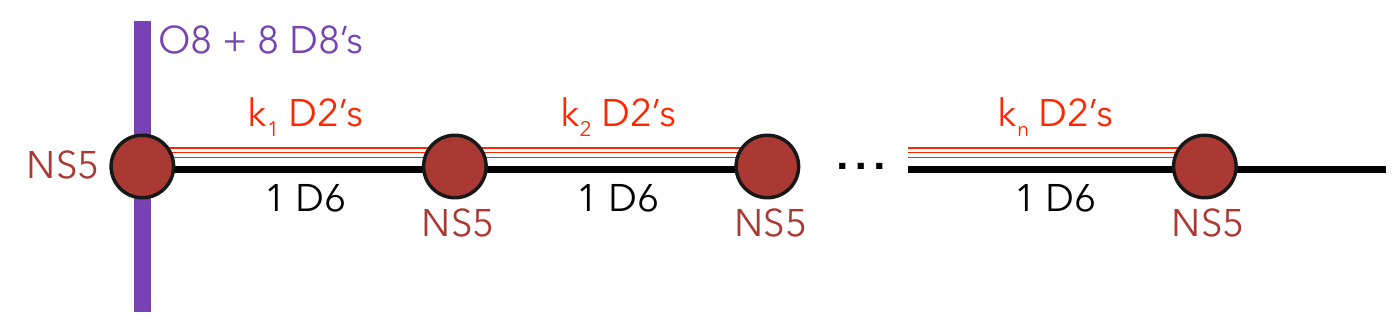}
  \caption{IIA brane system which uplifts to E-string theory on multiple M5-branes}
  \label{fig:multiple-ns5-brane}
\end{figure}

The IIA brane system in Section~\ref{sec:strings-from-branes} can be uplifted in the heterotic M-theory to an M9-brane intersecting $\mathbb{C}^2/\mathbb{Z}_{N}$, together with an M5-brane supported at the origin of $\mathbb{C}^2/\mathbb{Z}_{N}$ \cite{Hanany:1997gh,Aspinwall:1996vc,DelZotto:2014hpa}. At $N = 1$, the worldvolume theory on a D6-brane segment becomes E-string theory. This is why we expect the 2d gauge theory of Figure~\ref{fig:2d-quiver} to be a UV description for E-strings. One can also insert extra M5-branes at the origin of $\mathbb{C}^2/\mathbb{Z}_{N}$. It adds extra NS5-branes to the IIA brane set-up. Suppose there are $n$ M5-branes probing the M9-brane. In the IIA system, we have $n$ parallel NS5-branes separated from the O8-D8's along the semi-infinite $x^6$ direction. In between $i$'th and $(i+1)$th NS5-branes, there lies a single D6-brane segment on top of which $k_i$ D2-branes are placed. See Figure~\ref{fig:multiple-ns5-brane}. Finite D2-branes realize a chain of self-dual strings, which appears in E-string theory on many M5-branes.

\begin{figure}[h]
  \centering
  \vspace{0.1cm}
  \includegraphics[height=4.4cm]{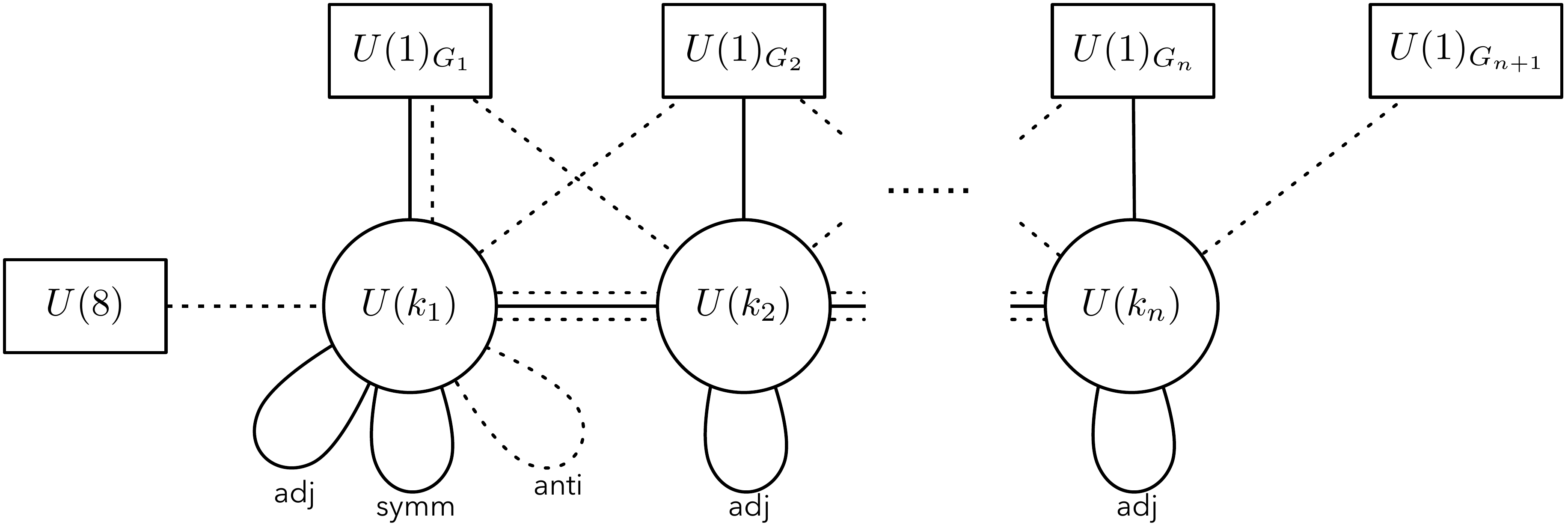}
  \caption{UV gauge theory for string chain on multiple M5-branes probing M9-brane}
  \label{fig:e-string-multiple-quiver}
\end{figure}

This system preserves the same $(0,4)$ SUSY $Q^{\dot{\alpha} A}_{-+}$ because extra NS5-brane does not break SUSYs. As each stack of $k_i$ D2-branes has a $U(k_i)$ gauge symmetry, the full quiver gauge theory has the $\prod_{a=1}^n U(k_a)$ gauge symmetry. It also sees a global symmetry $U(8)_{\rm D8} \times U(1)_A \times \prod_{a=1}^{n+1} U(1)_{G_a}$, where $U(8)_{\rm D8}$ and $U(1)_{G_a} $ correspond to gauge symmetries of the eight D8-branes and the $a$'th D6-brane. $U(1)_A$ is an extra global symmetry that comes from a flavor symmetry of underlying 6d SCFT. This theory contains various $(0,4)$ supermultiplets, being induced from open strings connecting D2's to other D2-, D6-, D8-branes. Below we present the table and Figure~\ref{fig:e-string-multiple-quiver} summarizing the field contents.
\begin{center}
  \begin{tabular}{c | c | cccc}
       Field & Type & $G$ rep. & $(\mathcal{Q}_{i-1},\mathcal{Q}_{i},\mathcal{Q}_{i+1})$ & $U(8)_{\rm D8}$ & $U(1)_A$\\
       \hline
       $(A_\mu^{(i)}, \lambda^{\dot{\alpha}A}_{(i)})$ & vector & \textbf{adj}$_{i}$ & $(0,0,0)$  & $-$ & 0\\
       $(a_{\alpha \dot{\beta}}^{(i)}, \lambda^{A}_{(i)\alpha})$ & hyper & \textbf{adj}$_i$ & $(0,0,0)$ & $-$ & 0\\
       $(q_{\dot{\alpha}}^{(i)}, \psi^A_{(i)})$ & hyper & $\mathbf{k}_i$ & $(0,-1,0)$ & $-$ & 0\\
       $(\rho_{(i)})$ & Fermi & $\mathbf{k}_{i}$ & $(0,0,-1)$ & $-$& 0\\
       $(\sigma_{(i)})$ & Fermi & $\overline{\mathbf{k}}_{i}$ & $(+1,0,0)$ & $-$& 0\\
       $(q_{A}^{(i)}, \psi^{\dot{\alpha}}_{(i)})$ & twisted hyper & $(\mathbf{k}_i, \overline{\mathbf{k}}_{i+1})$ & $(0,0,0)$ & $-$ & 0 \\
       $(\psi_{(i)})$ & Fermi & $(\mathbf{k}_i, \overline{\mathbf{k}}_{i+1})$ & $(0,0,0)$ & $-$ & 0\\ \hline
       $(\Xi_l)$ & Fermi & $\mathbf{k}_1$ & $(0,0,0)$ & $\mathbf{\overline{8}}$ & 0\\
       $(\varphi_{A}, \Phi^{\dot{\alpha}})$ & twisted hyper & $\mathbf{sym}_1$ & $(0,0,0)$ & $-$ & $+1$\\
       $(\Psi_{\alpha})$ & Fermi & \textbf{anti}$_1$ & $(0,0,0)$ & $-$ & $+1$ \\
       $(\psi)$ & Fermi & $\mathbf{k}_1$ & $\mathcal{Q}_1 = +1$ & $-$ & $+1$
   \end{tabular}    
\end{center}
$\mathcal{Q}_{i}$ denotes the $U(1)_{G_i}$ charge of a given field. There are $n$ sets of vector, hyper, and Fermi multiplets for each $U(k_i)$ gauge group, which are displayed on the first 4 lines. All $U(k_i)$'s except $i=1$ carry Fermi multiplets in the antifundamental representation denoted on the 5th line. Twisted hypermultiplets and Fermi multiplets on the 6th and 7th lines are bifundamentally charged under $U(k_i) \times U(k_{i+1})$.
All remainders are charged only under the $U(k_1)$ gauge group, as they are induced from open strings connecting the first $k_1$ D2-branes to adjacent D-branes.

Many $U(1)$ global symmetries of the above theory produce mixed anomalies with the Abelian part of gauge symmetry. If one denotes by $F_1$, $A$, $G_{1}$, $\cdots$ $G_{n}$, and $F_2$ respectively, the generators of $U(1)_{F_1} \subset U(8)$, $U(1)_A$, $U(1)_{G_{1}}$, $\cdots$, $U(1)_{G_{n}}$, $U(1)_{G_{n+1}}$, all non-zero mixed anomalies are given by
\begin{alignat}{4}
 &\text{Tr}\, (\gamma_3 \, T_{\rm U(1)}^{(1)}\, F_1) = 8, & \quad &\text{Tr}\, (\gamma_3 \, T_{\rm U(1)}^{(1)}\, A) = 3, & \quad &\text{Tr}\, (\gamma_3 \, T_{\rm U(1)}^{(1)}\, G_1) = - 3, & \quad &\text{Tr}\, (\gamma_3 \, T_{\rm U(1)}^{(1)}\, G_2) = 1 \nonumber\\
 \label{eq:anomaly-free-quiver}
 &\text{Tr}\, (\gamma_3 \, T_{\rm U(1)}^{(i)}\, G_{i-1}) = 1, & \quad &\text{Tr}\, (\gamma_3 \, T_{\rm U(1)}^{(i)}\, G_i) = -2, & \quad &\text{Tr}\, (\gamma_3 \, T_{\rm U(1)}^{(i)}\, G_{i+1}) = 1 & &\text{(for $2 \leq i \leq R$)}.
\end{alignat}
In a sensible quantum theory, all existing $U(1)$'s should be free of mixed anomalies. This implies that all global $U(1)$ symmetries must be locked into three $U(1)$ combinations in the quantum dynamics.
There also exists the fake $U(1)$ symmetry which is generated by $\sum_{i=1}^n T_{\rm U(1)}^{(i)} + \sum_{i=1}^n G_{i} + F_2 - A$. 

The elliptic genera for string chains in E-string theory on multiple M5-branes were studied in \cite{Gadde:2015tra}. \cite{Kim:2014dza, Gadde:2015tra} used a different IIA description of E-string theory, which was somewhat similar to Figure~\ref{fig:multiple-ns5-brane} while D6 and $\frac{1}{2}$NS5 were absent. Consider $n$ parallel M5-branes near a M9-brane
\begin{center}
  \vspace{-0.3cm}
  \begin{tabular}{c|ccccccccccc}
    & 0 & 1 & 2 & 3 & 4 & 5 & 6 & 7 & 8 & 9 & 10\\ \hline
    M5 & $\bullet$ & $\bullet$ & $\bullet$ & $\bullet$ & $\bullet$ & $\bullet$ & $-$ & $-$ &$-$ & $-$\\
    M9 & $\bullet$ & $\bullet$ & $\bullet$ & $\bullet$ & $\bullet$ & $\bullet$ & $-$ & $\bullet$ & $\bullet$ & $\bullet$ & $\bullet$\\
  \end{tabular}  
\end{center}
where the $789$ and $10$ directions were taken to be $\mathbb{R}^3 \times S^1_r$ with a radius $r$. In the 11d limit $r \rightarrow \infty$, E-string theory has the global $SO(4)_{789,10}$ symmetry which rotates four transverse directions. On the other hand, the IIA system in \cite{Kim:2014dza, Gadde:2015tra} was obtained by the $r \rightarrow 0$ limit which reduces the $x^{10}$ circle. Note that the $E_8$ Wilson line is required to reach the IIA configuration, replacing the M9 by the O8${}^-$ and 8 D8 branes. Even though the $x^{10}$ direction would be decompactified in the infrared regime of 2d gauge theory through the following relation $g \propto g_{s} \propto r \rightarrow \infty$, one could not capture the $SO(4)_{789,10}$ flavored spectrum due to a missing Cartan in the IIA set-up. In fact, \cite{Kim:2014dza, Gadde:2015tra} computed the $SO(3)_{789}$ flavored spectrum where $SO(3)_{789}$ is identified as the diagonal subgroup $SU(2)_D \subset SU(2)_L \times SU(2)_R = SO(4)_{789,10}$. For a single M5-brane, it does not matter because 5d $Sp(1)$ instanton calculus, which we explain below, shows that the BPS spectrum itself does not include $SU(2)_L$ charged states \cite{Kim:2012gu,Hwang:2014uwa}. $SU(2)_L$ charged states begin to appear at $n>1$. 5d $Sp(n)$ instanton calculus tells us that the BPS spectrum would be a product of $n$ $Sp(1)$ partition functions if not refined with the $SU(2)_L$ global symmetry \cite{Gadde:2015tra}, which happens because many BPS states cancel to 0 in the elliptic genus without $SU(2)_L$ refinement. This calls for a new description in which the $SO(4)_{789,10}$ symmetry is fully visible.

We can instead take the $789$ and $10$ directions to be Taub-NUT. M5-branes are supported at the origin of Taub-NUT, which locally looks like $\mathbb{R}^4$. The IIA brane set-up in Figure~\ref{fig:multiple-ns5-brane} is obtained in the limit which reduces the Taub-NUT circle. Again, it is required to turn on the $E_8$ Wilson line along a reduced circle, breaking $E_8$ to $SO(16)$ symmetry. Even though both IIA systems in \cite{Gadde:2015tra} and Figure~\ref{fig:multiple-ns5-brane} reach to the same E-string theory in the strong coupling regime, the IIA system in Figure~\ref{fig:multiple-ns5-brane} is more favorable because it realize both $SU(2)_L$ and $SU(2)_R$ Cartans as a part of flavor symmetry and the R-symmetry of 6d $(1,0)$ SUSY. It makes our 2d gauge theory, introduced in Figure~\ref{fig:e-string-multiple-quiver}, more useful when we study the fully refined spectrum of E-string theory. 

Let us explain how 5d $Sp(n)$ instanton calculus describes E-string theory. If one chooses one of $12345$ directions, shared by M5 and M9-branes, to be a circle $S^1_r$ while keeping $789$ and $10$ directions as non-compact $\mathbb{R}^4$, the IIA brane system can be obtained by reducing $S^1_r$ along which the $E_8$ Wilson line is turned on. M5-branes are mapped to D4-branes. M9-brane is replaced by 8 D8-branes on top of the O8${}^-$ plane, carrying the $SO(16)$ symmetry. The low energy dynamics of $n$ D4-branes is described by 5d $Sp(n)$ gauge theory. It has $8$ fundamental and $1$ antisymmetric hypermultiplets. A remarkable fact is that certain 5d gauge theories essentially capture 6d physics by incorporating non-perturbative objects called instantons \cite{Douglas:2010iu,Lambert:2010iw}. Recall that the low energy dynamics of 5d SYM instantons are UV-completed as D0-branes on D4-branes in string theory. As D0-branes are Kaluza-Klein momenta along the M-theory circle, one can interpret 5d SYM instantons as Kaluza-Klein momenta of a circle compactified 6d SCFT. This relation has been confirmed in certain examples. For example, the instanton partition function of 5d $\mathcal{N}=2$ SYM displays the tensor branch spectrum of 6d M-string theory \cite{Kim:2011mv,Haghighat:2013fc}. Along the same line, the instanton partition function of 5d $\mathcal{N}=1$ $Sp(1)$ gauge theory agrees with the tensor branch spectrum of 6d E-string theory on a single M5-brane \cite{Hwang:2014uwa,Kim:2014dza}. One can surely extend this relation to E-string theory on $n$ M5-branes, obtaining 5d $Sp(n)$ gauge theory on $n$ D4-branes. The precise relation between the 5d $Sp(n)$ instanton partition function and the 6d tensor branch spectrum is explained in \cite{Kim:2014dza,Gadde:2015tra}. We shall refer to \cite{Gadde:2015tra} when we compare the elliptic genera of the gauge theory in Figure~\ref{fig:e-string-multiple-quiver} with the 5d $Sp(n)$ instanton partition function.

Below we are going to compute the elliptic genera of 2d gauge theories in Figure~\ref{fig:2d-quiver} and Figure~\ref{fig:e-string-multiple-quiver}. The point of this computation is verifying the $E_8$ symmetry enhancement. For the latter case, we content ourselves with the sample computation at $n=2$ which will be compared with the 5d $Sp(2)$ instanton partition function \cite{Hwang:2014uwa}.

\paragraph{One M5-brane}

At $n=1$, we bring the $k=1$ and $k=2$ elliptic genera computed in Section~\ref{sec:IR-dynamics-symmetry}. One sets $\overline{a} = 0$ using the fake $U(1)$ symmetry, then the equation \eqref{eq:mixed-anomaly-elg} imposes $Y = \bar{y}^{-3}$.

Taking advantage of the $E_8 \longrightarrow SU(9)$ branching rules, e.g.,
\begin{align}
  \label{eq:su9-e8-branching-rule}
  \mathbf{248} & \longrightarrow \mathbf{84} + \overline{\mathbf{84}} + \mathbf{80}\\
  \mathbf{3875} & \nonumber\longrightarrow \mathbf{80} + \mathbf{240} + \overline{\mathbf{240}} + \mathbf{1050} + \overline{\mathbf{1050}} + \mathbf{1215}\\
  \mathbf{27000} & \nonumber\longrightarrow \mathbf{1}+\mathbf{80}+\mathbf{84}+\overline{\mathbf{84}}+\mathbf{1050}+\overline{\mathbf{1050}}+\mathbf{1215}+\mathbf{1944}\\&\qquad+\mathbf{2520}+\overline{\mathbf{2520}}+\mathbf{5346}+\overline{\mathbf{5346}}+\mathbf{5760} \nonumber\\
  \mathbf{30380} & \nonumber\longrightarrow \mathbf{1}+\mathbf{80}+\mathbf{84}+\overline{\mathbf{84}}+\mathbf{240}+\overline{\mathbf{240}}+\mathbf{1050}+\overline{\mathbf{1050}}+\mathbf{1215}+\mathbf{1540}+\overline{\mathbf{1540}}\\&\qquad+\mathbf{3402}+\overline{\mathbf{3402}}+\mathbf{5346}+\overline{\mathbf{5346}}+\mathbf{5760}\nonumber
\end{align}
the BPS spectrum at $k=1$ can be organized into $E_8$ representations as follows.
\begin{align}
   \frac{t}{(1-tu)(1-tu^{-1})} &\Big[q^{-1/2} + q^{1/2}\cdot (\chi_{1/2}^{\rm SU(2)}(t)\,\chi_{1/2}^{\rm SU(2)}(u) + \chi^{\rm E_8}_{\bf 248} ) +  q^{3/2} \cdot \Big(\chi_{1}^{\rm SU(2)}(t)\, \chi_{1}^{\rm SU(2)}(u)
   \\&+ \chi_{1/2}^{\rm SU(2)}(t)\, \chi_{1/2}^{\rm SU(2)}(u) (\chi^{\rm E_8}_{\bf 248} + 1)
   +  \chi^{\rm E_8}_{\bf 3875} + \chi^{\rm E_8}_{\bf 248} +2 \Big) + \mathcal{O}(q^{2})\Big] \nonumber
\end{align}
Similarly at $k=2$, we obtain
\begin{align}
   &\frac{t^2}{(1-tu)^2(1-tu^{-1})^2} \Bigg[  q^{-1} \cdot \frac{t \,\chi _{1/2}^{\text{SU}(2)}(t) }{(1+ tu^{-1}) (1+ t u)} + q^{0} \cdot (\chi^{\rm SU(2)}_{1/2}(t)\, \chi^{\rm SU(2)}_{1/2}(u) + \chi^{\rm E_8}_{\bf 248}) 
   \\& + q^{1} \cdot
    \Big\{ 3 \chi _{1/2}^{\text{SU}(2)}(t) \chi
   _1^{\text{SU}(2)}(u)+3 \chi _1^{\text{SU}(2)}(t) \chi_{1/2}^{\text{SU}(2)}(u)+2 \chi_{3/2}^{\text{SU}(2)}(t) \chi_1^{\text{SU}(2)}(u) 
   +\chi _1^{\text{SU}(2)}(t) \chi_{3/2}^{\text{SU}(2)}(u) \nonumber
   \\&\quad -\chi _{5/2}^{\text{SU}(2)}(t)\chi _1^{\text{SU}(2)}(u)+4 \chi_{1/2}^{\text{SU}(2)}(t)+4 \chi_{1/2}^{\text{SU}(2)}(u)
   +\chi_{3/2}^{\text{SU}(2)}(u) +\chi^{\rm E_8}_{\bf 248} \, \big(2 \chi _{1/2}^{\text{SU}(2)}(t)
   \chi_1^{\text{SU}(2)}(u) +3\chi _{1/2}^{\text{SU}(2)}(t)\nonumber\\
   &\quad +\chi _1^{\text{SU}(2)}(t) \chi _{3/2}^{\text{SU}(2)}(u) +\chi_{3/2}^{\text{SU}(2)}(t) \chi_2^{\text{SU}(2)}(u) +2 \chi _1^{\text{SU}(2)}(t) \chi_{1/2}^{\text{SU}(2)}(u) -\chi_2^{\text{SU}(2)}(t) \chi _{1/2}^{\text{SU}(2)}(u)\nonumber
   +3 \chi_{1/2}^{\text{SU}(2)}(u)\big) 
   \\&\quad + \chi^{\rm E_8}_{\bf 3875} \big( \chi _{1/2}^{\text{SU}(2)}(t) \chi_1^{\text{SU}(2)}(u)+\chi_{1/2}^{\text{SU}(2)}(t)-\chi_{3/2}^{\text{SU}(2)}(t)+\chi_{1/2}^{\text{SU}(2)}(u) \big) + \chi^{\rm E_8}_{\bf 27000} \, \chi_{1/2}^{\rm SU(2)}(t) + \chi^{\rm E_8}_{\bf 30380} \, \chi_{1/2}^{\rm SU(2)}(u) \Big\} \nonumber\\&
   \quad \times \frac{t}{(1+ tu^{-1}) (1+ t u)} + \mathcal{O}(q^2)\Bigg]. \nonumber
\end{align}
which shows the $E_8$ flavor symmetry. Note that the $U(1)_F$ fugacity $\overline{y}$ automatically disappears, as expected from the 5d $Sp(1)$ instanton calculus \cite{Kim:2012gu,Hwang:2014uwa}. These results agree with the elliptic genera computed from the 2d $O(k)$ gauge theory in Figure~\ref{fig:higgsed-ok}(b) \cite{Kim:2014dza}.

\paragraph{Two M5-branes}
We compute the elliptic genus at $k_1 = k_2 = 1$, which can be written as
\begin{align}
\oint d \phi_{1,2} \, \frac{\eta^6\, \theta_1(2 \epsilon_+)^2}{\theta_1(\epsilon_{1,2})^2} \frac{\theta_1 (\phi_{1,2}- a_{2,1}) \theta_1(\phi_2-m_9) \prod_{l=1}^8 \theta_1(\phi_1 - m_l)}{\eta^7\,\theta_1 (\epsilon_+ \pm (\phi_{1,2} - a_{1,2}))}\frac{\eta \, \theta_1(\phi_1 + a_1 + M)}{\theta_1 (-\epsilon_+ \pm (2 \phi_1 + M))} \frac{\theta_1(\epsilon_- \pm \phi_{12})}{\theta_1(-\epsilon_+ \pm \phi_{12})}.
\end{align}
$\{ a_{1},\ a_2,\ M, \ m_9\}$ are chemical potentials for global symmetries $\{ U(1)_{G_1},\ U(1)_{G_2},\ U(1)_A,\ U(1)_{G_3}\}$. We also denote a $U(1)_F \subset U(8)$ chemical potential by $\overline{m}$, related to $U(8)$ parameters $m_{1},\cdots,m_8$ by
\begin{alignat}{2}
    m_l &= \tilde{m}_l + \overline{m} && \quad \text{ where } \quad \textstyle\sum_{l=1}^{8} \tilde{m}_l = 0.
\end{alignat}
One first sets $a_1 = 0$ using the fictitious $U(1)$ symmetry, then locks various $U(1)$'s to others according to \eqref{eq:anomaly-free-quiver}. After all, two independent $U(1)$'s can remain which have no mixed anomaly. This imposes
\begin{align}
  \label{eq:anomaly-chemical-quiver}
    a_1 - 2a_2 + m_9 = 0, \quad 3M + 8\overline{m} -3a_1 + a_2 = 0.
\end{align}
We write down all contributing JK-residues, with the choice of $\mathfrak{n} = (+1,+1)$, in the following.
\begin{itemize}
  \item $(\phi_1, \phi_2) = (a_1 - \epsilon_+, a_2 - \epsilon_+)$.
  \begin{align}
      \frac{-\eta^{-6}}{\theta_1(\epsilon_{1,2})^2}\frac{\theta_1 (\epsilon_- \pm (a_1 - a_2))\, \theta_1 (a_2- m_9 - \epsilon_+) \prod_{l=1}^8 \theta_1 (a_1 -m_l- \epsilon_+)}{\theta_1 (2a_1 - 3 \epsilon_+ + M) } 
  \end{align}
  \item $(\phi_1, \phi_2) = (\frac{\epsilon_+ - M}{2} + \ell_I, a_2 - \epsilon_+)$ for  $\ell = \{0, \frac{1}{2},\frac{1+\tau}{2},\frac{\tau}{2}\}$ \quad ($I=1,2,3,4$)
  \begin{align}
      \frac{\eta^{-6} \theta_1 (\epsilon_+ + a_1 - a_2) \theta_1 (\epsilon_+ - a_2 + m_9)}{2\theta_1(\epsilon_{1,2})^2} 
      \sum_{I=1}^4 \frac{s_I \theta_I(\frac{M-\epsilon_+}{2} -\epsilon_{1,2} + a_2) \prod_{l=1}^8 \theta_I(\frac{M-\epsilon_+}{2}+m_l)}{\theta_I(\frac{M-3 \epsilon_+}{2}+a_1) \theta_I(\frac{M - 5 \epsilon_+}{2}+a_2)}
  \end{align}
  where the sign factor $s_I$ is defined to be $s_1 = -1$ and $s_{2,3,4} = +1$.
  \item $(\phi_1, \phi_2) = (\frac{\epsilon_+ - M}{2} + \ell_I, \frac{3\epsilon_+ - M}{2} + \ell_I)$.
  \begin{align}
      \frac{-\eta^{-6}}{2\theta_1(\epsilon_{1,2})} \sum_{I=1}^4 \frac{s_I  \theta_I(\frac{M-3 \epsilon_+}{2} +m_9) \prod_{l=1}^8 \theta_I(\frac{M-\epsilon_+}{2}+ m_l)}{\theta_I(\frac{M-5 \epsilon_+}{2}+a_2)}.
  \end{align}
\end{itemize}

The condition \eqref{eq:anomaly-chemical-quiver} is translated into $a_2 = \frac{1}{2} m_9$ and $M = -\frac{8}{3}\overline{m}-\frac{1}{6} m_9$. We also rearrange nine independent variables $\tilde{m}_1$, $\cdots$, $\tilde{m}_7$, $\overline{m}$, $m_9$ into
\begin{align}
    \tilde{m}_l = m_l, \quad \tilde{m}_9 = 2 m_a,\quad \overline{m} = m_a + 3 \mathfrak{m}
\end{align}
in order that the $SU(9) \times SU(2)_L$ global symmetry becomes manifest. $m_a$ and $\mathfrak{m}$ are chemical potentials for $SU(2)_L$ and $U(1)$ symmetries, where the $U(1)$ appears in the $SU(9)$ symmetry enhancement from $SU(8) \times U(1)$. We denote the corresponding fugacities by $y_a \equiv e^{2\pi i m_a}$ and $\mathfrak{y} \equiv e^{2\pi i \mathfrak{m}}$. Organizing the spectrum, we use the $SU(9) \longrightarrow SU(8)\times U(1)$ branching rules, e.g.,
\begin{align*}
    \mathbf{80} &\longrightarrow \mathbf{1}_0 + \mathbf{8}_{+3} + \overline{\mathbf{8}}_{-3} + \mathbf{63}_0\\
    \mathbf{84} &\longrightarrow \mathbf{28}_{-2} + \mathbf{56}_{+1}\\
    \overline{\mathbf{84}} &\longrightarrow \overline{\mathbf{28}}_{+2} + \overline{\mathbf{56}}_{-1}.
\end{align*}
After the $SU(9)$ flavor symmetry enhancement, we observe that the $SU(9)$ further enhances to $E_8$. If one utilizes the $E_8 \longrightarrow SU(9)$ branching rules \eqref{eq:su9-e8-branching-rule}, $q$-expansion of the elliptic genus can be written in terms of $E_8$ characters.
\begin{align}
    &\frac{t^2}{(1-tu)^2(1-tu^{-1})^2} \Bigg[q^{-1/2}\,  \big\{\chi^{\rm SU(2)}_{1/2} (y_a) - \chi^{\rm SU(2)}_{1/2} (u)\big\}+q^{1/2}\,  \Big\{
    2 \chi _{1/2}^{\text{SU}(2)}(t) \chi _{1/2}^{\text{SU}(2)}(u) \chi _{1/2}^{\text{SU}(2)}(y_a)\\
    & -\chi _{1/2}^{\text{SU}(2)}(u) \chi _1^{\text{SU}(2)}(y_a)+\chi _1^{\text{SU}(2)}(u) \chi _{1/2}^{\text{SU}(2)}(y_a)+\chi _{1/2}^{\text{SU}(2)}\left(y_a\right)-2 \chi _{1/2}^{\text{SU}(2)}(t) \chi _1^{\text{SU}(2)}(u)-2 \chi _{1/2}^{\text{SU}(2)}(t)\nonumber\\
    &-\chi _{1/2}^{\text{SU}(2)}(u) + 
    \chi^{\rm E_8}_{\bf 248}\big(-\chi _{1/2}^{\text{SU}(2)}(u)+\chi _{1/2}^{\text{SU}(2)}(y_a)\big)
    +q^{3/2} \Big\{ \chi^{\rm E_8}_{\bf 3875} \big( 
    \chi^{\rm SU(2)}_{1/2} (y_a) - \chi^{\rm SU(2)}_{1/2} (u)\big) + \chi^{\rm E_8}_{\bf 248} \big(
    -2 \chi _{1/2}^{\text{SU}(2)}(t)\nonumber\\&
    -2 \chi _{1/2}^{\text{SU}(2)}(u)
    -2 \chi _{1/2}^{\text{SU}(2)}(t) \chi _1^{\text{SU}(2)}(u) \chi _1^{\text{SU}(2)}\left(y_a\right)+2 \chi _{1/2}^{\text{SU}(2)}(t) \chi _{1/2}^{\text{SU}(2)}(u) \chi _{1/2}^{\text{SU}(2)}\left(y_a\right)+\chi _{3/2}^{\text{SU}(2)}\left(y_a\right) \nonumber
    \\&+2 \chi _{1/2}^{\text{SU}(2)}(t) \chi _{3/2}^{\text{SU}(2)}(u) \chi _{1/2}^{\text{SU}(2)}\left(y_a\right)+3 \chi _1^{\text{SU}(2)}(t) \chi _1^{\text{SU}(2)}(u) \chi _{1/2}^{\text{SU}(2)}\left(y_a\right)+\chi _1^{\text{SU}(2)}(t) \chi _{1/2}^{\text{SU}(2)}\left(y_a\right)\nonumber\\&
    -\chi _{1/2}^{\text{SU}(2)}(u) \chi _1^{\text{SU}(2)}\left(y_a\right)+7 \chi _1^{\text{SU}(2)}(u) \chi _{1/2}^{\text{SU}(2)}\left(y_a\right)+2 \chi _{1/2}^{\text{SU}(2)}\left(y_a\right)-2 \chi _{1/2}^{\text{SU}(2)}(t) \chi _1^{\text{SU}(2)}(u)\nonumber
    \\&-4 \chi _1^{\text{SU}(2)}(t) \chi _{1/2}^{\text{SU}(2)}(u)-3 \chi _1^{\text{SU}(2)}(t) \chi _{3/2}^{\text{SU}(2)}(u) \big) 
    -2 \chi _{1/2}^{\text{SU}(2)}(t) \chi _1^{\text{SU}(2)}(u) \chi _1^{\text{SU}(2)}\left(y_a\right) -4 \chi _{1/2}^{\text{SU}(2)}(t)\nonumber\\
    &+6 \chi _{1/2}^{\text{SU}(2)}(t) \chi _{1/2}^{\text{SU}(2)}(u) \chi _{1/2}^{\text{SU}(2)}\left(y_a\right)+2 \chi _{1/2}^{\text{SU}(2)}(t) \chi _{3/2}^{\text{SU}(2)}(u) \chi _{1/2}^{\text{SU}(2)}\left(y_a\right)+3 \chi _1^{\text{SU}(2)}(t) \chi _1^{\text{SU}(2)}(u) \chi _{1/2}^{\text{SU}(2)}\left(y_a\right)\nonumber\\&
    -2 \chi _{1/2}^{\text{SU}(2)}(t) \chi _1^{\text{SU}(2)}\left(y_a\right)+\chi _1^{\text{SU}(2)}(t) \chi _{1/2}^{\text{SU}(2)}\left(y_a\right)-2 \chi _{1/2}^{\text{SU}(2)}(u) \chi _1^{\text{SU}(2)}\left(y_a\right)+3 \chi _1^{\text{SU}(2)}(u) \chi _{1/2}^{\text{SU}(2)}\left(y_a\right)\nonumber\\&
    +6 \chi _{1/2}^{\text{SU}(2)}\left(y_a\right)+\chi _{3/2}^{\text{SU}(2)}\left(y_a\right)-4 \chi _{1/2}^{\text{SU}(2)}(t) \chi _1^{\text{SU}(2)}(u)-4 \chi _1^{\text{SU}(2)}(t) \chi _{1/2}^{\text{SU}(2)}(u)-3 \chi _1^{\text{SU}(2)}(t) \chi _{3/2}^{\text{SU}(2)}(u)\nonumber\\&
    -7 \chi _{1/2}^{\text{SU}(2)}(u)-2 \chi _{3/2}^{\text{SU}(2)}(u)\Big\} + \mathcal{O}(q^2)\Bigg]
\end{align}
We checked that it agrees with 5d $Sp(2)$ instanton partition function \cite{Hwang:2014uwa}.

\section{Concluding remarks}
\label{sec:conclusion}

In this paper, we studied those 6d $SU(3)$ and $SU(2)$ SCFTs which can be Higgsed to M-string theory and E-string theory. We constructed the anomaly-free 2d $(0,4)$ gauge theories for self-dual strings, inspired from the ADHM construction. This issue is   closely related to finding a proper UV completion of the non-linear sigma model which governs the low energy dynamics of instanton string solitons in 6d effective SYM. 
Our 2d consistent gauge theories capture the global symmetries of self-dual strings in the IR limit, implying that they provide a UV description for 6d instanton strings. 

Similar to the E-string case, it would be interesting to check our results for $SU(2)$ and $SU(3)$  independently, for example,  by considering the instanton partition function of the dual 5d theories obtained after the circle compactification~\cite{Hayashi:2015fsa,Yonekura:2015ksa}.

Our analysis can be applied to other 6d SCFTs and little string theories, which contains 6d supersymmetric gauge theory as effective descriptions in tensor branch. The 2d elliptic genus calculation can be an interesting starting point for understanding the index function of the 6d $(1,0)$ SCFTs on $S^1\times S^5$.  We hope to come back to these issues in near future.

\vspace{0.8cm}

\noindent{\bf\large Acknowledgements}

\noindent
The work of SK is supported in part by the National Research Foundation of Korea (NRF) Grant NRF-2015R1A2A2A01003124. The work of KL is supported in part by the National Research Foundation of Korea (NRF) Grants No. 2006-0093850.


\begin{thebibliography}{10}

\bibitem{Witten:1995zh}
E.~Witten, ``{Some comments on string dynamics},'' in {\em Future perspectives
  in string theory. Proceedings,Conference, Strings'95, Los Angeles, USA, March
  13-18,1995}.
\newblock 1995.
\newblock
\href{http://arxiv.org/abs/hep-th/9507121}{{\ttfamily arXiv:hep-th/9507121
  [hep-th]}}.
\newblock

\bibitem{Strominger:1995ac}
A.~Strominger, ``{Open p-branes},''
  \href{http://dx.doi.org/10.1016/0370-2693(96)00712-5}{{\em Phys. Lett.}
  {\bfseries B383} (1996) 44--47},
\href{http://arxiv.org/abs/hep-th/9512059}{{\ttfamily arXiv:hep-th/9512059
  [hep-th]}}.

\bibitem{Morrison:1996na}
D.~R. Morrison and C.~Vafa, ``{Compactifications of F theory on Calabi-Yau
  threefolds. 1},'' \href{http://dx.doi.org/10.1016/0550-3213(96)00242-8}{{\em
  Nucl. Phys.} {\bfseries B473} (1996) 74--92},
\href{http://arxiv.org/abs/hep-th/9602114}{{\ttfamily arXiv:hep-th/9602114
  [hep-th]}}.

\bibitem{Seiberg:1996vs}
N.~Seiberg and E.~Witten, ``{Comments on string dynamics in six-dimensions},''
  \href{http://dx.doi.org/10.1016/0550-3213(96)00189-7}{{\em Nucl. Phys.}
  {\bfseries B471} (1996) 121--134},
\href{http://arxiv.org/abs/hep-th/9603003}{{\ttfamily arXiv:hep-th/9603003
  [hep-th]}}.

\bibitem{Witten:1996qb}
E.~Witten, ``{Phase transitions in M theory and F theory},''
  \href{http://dx.doi.org/10.1016/0550-3213(96)00212-X}{{\em Nucl. Phys.}
  {\bfseries B471} (1996) 195--216},
\href{http://arxiv.org/abs/hep-th/9603150}{{\ttfamily arXiv:hep-th/9603150
  [hep-th]}}.

\bibitem{Morrison:1996pp}
D.~R. Morrison and C.~Vafa, ``{Compactifications of F theory on Calabi-Yau
  threefolds. 2.},'' \href{http://dx.doi.org/10.1016/0550-3213(96)00369-0}{{\em
  Nucl. Phys.} {\bfseries B476} (1996) 437--469},
\href{http://arxiv.org/abs/hep-th/9603161}{{\ttfamily arXiv:hep-th/9603161
  [hep-th]}}.

\bibitem{Blum:1997mm}
J.~D. Blum and K.~A. Intriligator, ``{New phases of string theory and 6-D RG
  fixed points via branes at orbifold singularities},''
  \href{http://dx.doi.org/10.1016/S0550-3213(97)00449-5}{{\em Nucl. Phys.}
  {\bfseries B506} (1997) 199--222},
\href{http://arxiv.org/abs/hep-th/9705044}{{\ttfamily arXiv:hep-th/9705044
  [hep-th]}}.

\bibitem{Brunner:1997gf}
I.~Brunner and A.~Karch, ``{Branes at orbifolds versus Hanany Witten in
  six-dimensions},''
  \href{http://dx.doi.org/10.1088/1126-6708/1998/03/003}{{\em JHEP} {\bfseries
  03} (1998) 003},
\href{http://arxiv.org/abs/hep-th/9712143}{{\ttfamily arXiv:hep-th/9712143
  [hep-th]}}.

\bibitem{Hanany:1997gh}
A.~Hanany and A.~Zaffaroni, ``{Branes and six-dimensional supersymmetric
  theories},'' \href{http://dx.doi.org/10.1016/S0550-3213(98)00355-1}{{\em
  Nucl. Phys.} {\bfseries B529} (1998) 180--206},
\href{http://arxiv.org/abs/hep-th/9712145}{{\ttfamily arXiv:hep-th/9712145
  [hep-th]}}.

\bibitem{DelZotto:2014hpa}
M.~Del~Zotto, J.~J. Heckman, A.~Tomasiello, and C.~Vafa, ``{6d Conformal
  Matter},'' \href{http://dx.doi.org/10.1007/JHEP02(2015)054}{{\em JHEP}
  {\bfseries 02} (2015) 054},
\href{http://arxiv.org/abs/1407.6359}{{\ttfamily arXiv:1407.6359 [hep-th]}}.

\bibitem{Morrison:2012np}
D.~R. Morrison and W.~Taylor, ``{Classifying bases for 6D F-theory models},''
  \href{http://dx.doi.org/10.2478/s11534-012-0065-4}{{\em Central Eur. J.
  Phys.} {\bfseries 10} (2012) 1072--1088},
\href{http://arxiv.org/abs/1201.1943}{{\ttfamily arXiv:1201.1943 [hep-th]}}.

\bibitem{Heckman:2013pva}
J.~J. Heckman, D.~R. Morrison, and C.~Vafa, ``{On the Classification of 6D
  SCFTs and Generalized ADE Orbifolds},''
  \href{http://dx.doi.org/10.1007/JHEP06(2015)017,
  10.1007/JHEP05(2014)028}{{\em JHEP} {\bfseries 05} (2014) 028},
  \href{http://arxiv.org/abs/1312.5746}{{\ttfamily arXiv:1312.5746 [hep-th]}}.
[Erratum: JHEP06,017(2015)].

\bibitem{Heckman:2015bfa}
J.~J. Heckman, D.~R. Morrison, T.~Rudelius, and C.~Vafa, ``{Atomic
  Classification of 6D SCFTs},''
  \href{http://dx.doi.org/10.1002/prop.201500024}{{\em Fortsch. Phys.}
  {\bfseries 63} (2015) 468--530},
\href{http://arxiv.org/abs/1502.05405}{{\ttfamily arXiv:1502.05405 [hep-th]}}.

\bibitem{Bhardwaj:2015xxa}
L.~Bhardwaj, ``{Classification of 6d N=(1,0) gauge theories},''
\href{http://arxiv.org/abs/1502.06594}{{\ttfamily arXiv:1502.06594 [hep-th]}}.

\bibitem{Klemm:1996bj}
A.~Klemm, W.~Lerche, P.~Mayr, C.~Vafa, and N.~P. Warner, ``{Selfdual strings
  and N=2 supersymmetric field theory},''
  \href{http://dx.doi.org/10.1016/0550-3213(96)00353-7}{{\em Nucl. Phys.}
  {\bfseries B477} (1996) 746--766},
\href{http://arxiv.org/abs/hep-th/9604034}{{\ttfamily arXiv:hep-th/9604034
  [hep-th]}}.

\bibitem{Schwarz:1996pi}
J.~H. Schwarz, ``{Selfdual superstring in six-dimensions},''
\href{http://arxiv.org/abs/hep-th/9604171}{{\ttfamily arXiv:hep-th/9604171
  [hep-th]}}.

\bibitem{Haghighat:2013fc}
B.~Haghighat, A.~Iqbal, C.~Kozcaz, G.~Lockhart, and C.~Vafa, ``{M-Strings},''
  {\em arXiv.org} no.~2, (May, 2013) 779--842,
  \href{http://arxiv.org/abs/1305.6322v2}{{\ttfamily 1305.6322v2}}.

\bibitem{Haghighat:2013fz}
B.~Haghighat, C.~Kozcaz, G.~Lockhart, and C.~Vafa, ``{On orbifolds of
  M-Strings},'' {\em arXiv.org} no.~4, (Oct., 2013) 046003,
  \href{http://arxiv.org/abs/1310.1185v1}{{\ttfamily 1310.1185v1}}.

\bibitem{Ganor:1996mu}
O.~J. Ganor and A.~Hanany, ``{Small E(8) instantons and tensionless noncritical
  strings},'' \href{http://dx.doi.org/10.1016/0550-3213(96)00243-X}{{\em Nucl.
  Phys.} {\bfseries B474} (1996) 122--140},
\href{http://arxiv.org/abs/hep-th/9602120}{{\ttfamily arXiv:hep-th/9602120
  [hep-th]}}.

\bibitem{Klemm:1996dt}
A.~Klemm, P.~Mayr, and C.~Vafa, ``{BPS States of Exceptional Non-Critical
  Strings},'' {\em arXiv.org} (July, 1996) ,
  \href{http://arxiv.org/abs/hep-th/9607139v2}{{\ttfamily hep-th/9607139v2}}.

\bibitem{Minahan:1998vr}
J.~A. Minahan, D.~Nemeschansky, C.~Vafa, and N.~P. Warner, ``{E strings and N=4
  topological Yang-Mills theories},''
  \href{http://dx.doi.org/10.1016/S0550-3213(98)00426-X}{{\em Nucl. Phys.}
  {\bfseries B527} (1998) 581--623},
\href{http://arxiv.org/abs/hep-th/9802168}{{\ttfamily arXiv:hep-th/9802168
  [hep-th]}}.

\bibitem{Eguchi:2002fc}
T.~Eguchi and K.~Sakai, ``{Seiberg-Witten curve for the E string theory},''
  \href{http://dx.doi.org/10.1088/1126-6708/2002/05/058}{{\em JHEP} {\bfseries
  05} (2002) 058},
\href{http://arxiv.org/abs/hep-th/0203025}{{\ttfamily arXiv:hep-th/0203025
  [hep-th]}}.

\bibitem{Kim:2014dza}
J.~Kim, S.~Kim, K.~Lee, J.~Park, and C.~Vafa, ``{Elliptic Genus of
  E-strings},''
\href{http://arxiv.org/abs/1411.2324}{{\ttfamily arXiv:1411.2324 [hep-th]}}.

\bibitem{Danielsson:1997kt}
U.~H. Danielsson, G.~Ferretti, J.~Kalkkinen, and P.~Stjernberg, ``{Notes on
  supersymmetric gauge theories in five-dimensions and six-dimensions},''
  \href{http://dx.doi.org/10.1016/S0370-2693(97)00645-X}{{\em Phys. Lett.}
  {\bfseries B405} (1997) 265--270},
\href{http://arxiv.org/abs/hep-th/9703098}{{\ttfamily arXiv:hep-th/9703098
  [hep-th]}}.

\bibitem{Bershadsky:1997sb}
M.~Bershadsky and C.~Vafa, ``{Global anomalies and geometric engineering of
  critical theories in six-dimensions},''
\href{http://arxiv.org/abs/hep-th/9703167}{{\ttfamily arXiv:hep-th/9703167
  [hep-th]}}.

\bibitem{Gadde:2015tra}
A.~Gadde, B.~Haghighat, J.~Kim, S.~Kim, G.~Lockhart, and C.~Vafa, ``{6d String
  Chains},''
\href{http://arxiv.org/abs/1504.04614}{{\ttfamily arXiv:1504.04614 [hep-th]}}.

\bibitem{Berkooz:1996iz}
M.~Berkooz, R.~G. Leigh, J.~Polchinski, J.~H. Schwarz, N.~Seiberg, and
  E.~Witten, ``{Anomalies, dualities, and topology of D = 6 N=1 superstring
  vacua},'' \href{http://dx.doi.org/10.1016/0550-3213(96)00339-2}{{\em Nucl.
  Phys.} {\bfseries B475} (1996) 115--148},
\href{http://arxiv.org/abs/hep-th/9605184}{{\ttfamily arXiv:hep-th/9605184
  [hep-th]}}.

\bibitem{Seiberg:1996qx}
N.~Seiberg, ``{Nontrivial fixed points of the renormalization group in
  six-dimensions},''
  \href{http://dx.doi.org/10.1016/S0370-2693(96)01424-4}{{\em Phys. Lett.}
  {\bfseries B390} (1997) 169--171},
\href{http://arxiv.org/abs/hep-th/9609161}{{\ttfamily arXiv:hep-th/9609161
  [hep-th]}}.

\bibitem{Atiyah:1978ri}
M.~F. Atiyah, N.~J. Hitchin, V.~G. Drinfeld, and {\relax Yu}.~I. Manin,
  ``{Construction of Instantons},''
\href{http://dx.doi.org/10.1016/0375-9601(78)90141-X}{{\em Phys. Lett.}
  {\bfseries A65} (1978) 185--187}.

\bibitem{Witten:1994tz}
E.~Witten, ``{Sigma models and the ADHM construction of instantons},''
  \href{http://dx.doi.org/10.1016/0393-0440(94)00047-8}{{\em J. Geom. Phys.}
  {\bfseries 15} (1995) 215--226},
\href{http://arxiv.org/abs/hep-th/9410052}{{\ttfamily arXiv:hep-th/9410052
  [hep-th]}}.

\bibitem{Douglas:1996sw}
M.~R. Douglas and G.~W. Moore, ``{D-branes, quivers, and ALE instantons},''
\href{http://arxiv.org/abs/hep-th/9603167}{{\ttfamily arXiv:hep-th/9603167
  [hep-th]}}.

\bibitem{Kim:2011mv}
H.-C. Kim, S.~Kim, E.~Koh, K.~Lee, and S.~Lee, ``{On instantons as Kaluza-Klein
  modes of M5-branes},'' \href{http://dx.doi.org/10.1007/JHEP12(2011)031}{{\em
  JHEP} {\bfseries 12} (2011) 031},
\href{http://arxiv.org/abs/1110.2175}{{\ttfamily arXiv:1110.2175 [hep-th]}}.

\bibitem{Shadchin:2005mx}
S.~Shadchin, {\em {On certain aspects of string theory/gauge theory
  correspondence}}.
\newblock PhD thesis, Orsay, LPT, 2005.
\newblock
\href{http://arxiv.org/abs/hep-th/0502180}{{\ttfamily arXiv:hep-th/0502180
  [hep-th]}}.
\newblock

\bibitem{Hwang:2014uwa}
C.~Hwang, J.~Kim, S.~Kim, and J.~Park, ``{General instanton counting and 5d
  SCFT},'' \href{http://dx.doi.org/10.1007/JHEP07(2015)063}{{\em JHEP}
  {\bfseries 07} (2015) 063},
\href{http://arxiv.org/abs/1406.6793}{{\ttfamily arXiv:1406.6793 [hep-th]}}.

\bibitem{Benini:2013xpa}
F.~Benini, R.~Eager, K.~Hori, and Y.~Tachikawa, ``{Elliptic Genera of 2d
  ${\mathcal{N}}$ = 2 Gauge Theories},''
  \href{http://dx.doi.org/10.1007/s00220-014-2210-y}{{\em Commun. Math. Phys.}
  {\bfseries 333} no.~3, (2015) 1241--1286},
\href{http://arxiv.org/abs/1308.4896}{{\ttfamily arXiv:1308.4896 [hep-th]}}.

\bibitem{Tong:2014yna}
D.~Tong, ``{The holographic dual of $AdS_{3} \times S^{3} \times S^{3} \times
  S^{1}$},'' \href{http://dx.doi.org/10.1007/JHEP04(2014)193}{{\em JHEP}
  {\bfseries 04} (2014) 193},
\href{http://arxiv.org/abs/1402.5135}{{\ttfamily arXiv:1402.5135 [hep-th]}}.

\bibitem{Benini:2013nda}
F.~Benini, R.~Eager, K.~Hori, and Y.~Tachikawa, ``{Elliptic genera of
  two-dimensional N=2 gauge theories with rank-one gauge groups},''
  \href{http://dx.doi.org/10.1007/s11005-013-0673-y}{{\em Lett. Math. Phys.}
  {\bfseries 104} (2014) 465--493},
\href{http://arxiv.org/abs/1305.0533}{{\ttfamily arXiv:1305.0533 [hep-th]}}.

\bibitem{Aspinwall:1996vc}
P.~S. Aspinwall, ``{Point - like instantons and the spin (32) / Z(2) heterotic
  string},'' \href{http://dx.doi.org/10.1016/S0550-3213(97)00232-0}{{\em Nucl.
  Phys.} {\bfseries B496} (1997) 149--176},
\href{http://arxiv.org/abs/hep-th/9612108}{{\ttfamily arXiv:hep-th/9612108
  [hep-th]}}.

\bibitem{Kim:2012gu}
H.-C. Kim, S.-S. Kim, and K.~Lee, ``{5-dim Superconformal Index with Enhanced
  En Global Symmetry},'' \href{http://dx.doi.org/10.1007/JHEP10(2012)142}{{\em
  JHEP} {\bfseries 10} (2012) 142},
\href{http://arxiv.org/abs/1206.6781}{{\ttfamily arXiv:1206.6781 [hep-th]}}.

\bibitem{Douglas:2010iu}
M.~R. Douglas, ``{On D=5 super Yang-Mills theory and (2,0) theory},''
  \href{http://dx.doi.org/10.1007/JHEP02(2011)011}{{\em JHEP} {\bfseries 02}
  (2011) 011},
\href{http://arxiv.org/abs/1012.2880}{{\ttfamily arXiv:1012.2880 [hep-th]}}.

\bibitem{Lambert:2010iw}
N.~Lambert, C.~Papageorgakis, and M.~Schmidt-Sommerfeld, ``{M5-Branes,
  D4-Branes and Quantum 5D super-Yang-Mills},''
  \href{http://dx.doi.org/10.1007/JHEP01(2011)083}{{\em JHEP} {\bfseries 01}
  (2011) 083},
\href{http://arxiv.org/abs/1012.2882}{{\ttfamily arXiv:1012.2882 [hep-th]}}.

\bibitem{Hayashi:2015fsa}
H.~Hayashi, S.-S. Kim, K.~Lee, M.~Taki, and F.~Yagi, ``{A new 5d description of
  6d D-type minimal conformal matter},''
  \href{http://dx.doi.org/10.1007/JHEP08(2015)097}{{\em JHEP} {\bfseries 08}
  (2015) 097},
\href{http://arxiv.org/abs/1505.04439}{{\ttfamily arXiv:1505.04439 [hep-th]}}.

\bibitem{Yonekura:2015ksa}
K.~Yonekura, ``{Instanton operators and symmetry enhancement in 5d
  supersymmetric quiver gauge theories},''
  \href{http://dx.doi.org/10.1007/JHEP07(2015)167}{{\em JHEP} {\bfseries 07}
  (2015) 167},
\href{http://arxiv.org/abs/1505.04743}{{\ttfamily arXiv:1505.04743 [hep-th]}}.

\end{thebibliography}

\pagebreak
\providecommand{\href}[2]{#2}\begingroup\raggedright\endgroup

\end{document}